\documentclass[12pt]{iopart}

\expandafter\let\csname equation*\endcsname\relax

\expandafter\let\csname endequation*\endcsname\relax

\usepackage{etoolbox}
\usepackage{amsmath}
\usepackage{amsfonts}
\usepackage{graphicx}
\usepackage{soul}
\usepackage{float} % for [H] in figures
\usepackage[dvipsnames]{xcolor}
\usepackage{tikz}
\usetikzlibrary{decorations.markings}
\usetikzlibrary{math}
\usetikzlibrary{backgrounds}
\usepackage[compat=1.1.0]{tikz-feynman}
\usepackage{ifthen}
\usepackage{xargs}
\usepackage[colorinlistoftodos]{todonotes}
\usepackage{hyperref}
\usepackage[numbers,square, sort&compress]{natbib}

\usepackage{mathtools}
\usepackage{mathdots}
\usepackage{amssymb}
\usepackage{cancel}
\usepackage[normalem]{ulem}

\hypersetup{
	colorlinks=true, %set true if you want colored links
	linktoc=all,     %set to all if you want both sections and subsections linked
	linkcolor=blue,
	urlcolor = blue,
}

\newcounter{todoListItems}
 \newcommand{\todoTrans}[2][ ]{
   \addtocounter{todoListItems}{1}
   \todo[%
     caption={\protect\hypertarget{todo\thetodoListItems}{}#2},
#1] {
#2 \hfill
     \hyperlink{todo\thetodoListItems}{$\uparrow$}
   }
}
\newcommandx{\change}[2][1=]{\todoTrans[linecolor=red,backgroundcolor=red!25,bordercolor=red,#1]{#2}}
\newcommandx{\unsure}[2][1=]{\todoTrans[linecolor=Magenta,backgroundcolor=Magenta!25,bordercolor=Magenta,#1]{#2}}
\newcommandx{\expandon}[2][1=]{\todoTrans[linecolor=ForestGreen,backgroundcolor=ForestGreen!25,bordercolor=ForestGreen,#1]{#2}}
\newcommandx{\add}[2][1=]{\todoTrans[linecolor=SkyBlue,backgroundcolor=SkyBlue!25,bordercolor=SkyBlue,#1]{#2}}
\makeatletter
\def\@mkboth#1#2{}
\newlength\appendixwidth
\preto\appendix{\addtocontents{toc}{\protect\patchl@section}}
\newcommand{\patchl@section}{%
  \settowidth{\appendixwidth}{\textbf{Appendix }}%
  \addtolength{\appendixwidth}{1.5em}%
  \patchcmd{\l@section}{1.5em}{\appendixwidth}{}{\ddt}%
}

\newcommand{\para}[1]{\left( #1 \right)}
\newcommand{\brac}[1]{\left[ #1 \right]}

\newcommand{\abs}[1]{\left\vert #1 \right\vert}

\newcommand{\braket}[1]{\left<#1\right>}
\newcommand{\bra}[1]{\left<#1\right|}
\newcommand{\ket}[1]{\left|#1\right>}
\newcommand{\ketbra}[2]{\left\vert#1\middle>\middle<#2\right\vert}
\newcommand{\braOket}[3]{\left<#1\middle\vert#2\middle\vert#3\right>}

\newcommand{\A}{\mathcal{A}}
\newcommand{\B}{\mathcal{B}}
\newcommand{\E}{\mathcal{E}}
\newcommand{\F}{\mathcal{F}}
\newcommand{\T}{\mathcal{T}}
\newcommand{\RNn}{\mathcal{R}}
\newcommand{\C}{\mathbb{C}}
\newcommand{\R}{\mathbb{R}}
\newcommand{\Z}{\mathbb{Z}}
\renewcommand{\H}{\mathcal{H}}
\renewcommand{\L}{\mathcal{L}}
\newcommand{\eli}{\text{...}}
\newcommand{\rtwc}[4]{\vartheta\begin{bmatrix}#1 \\ #2 \end{bmatrix}\left(#3\middle|#4\right)}
\newcommand{\rt}[2]{\vartheta\left(#1\middle|#2\right)}

\let\Im\relax
\DeclareMathOperator{\Im}{Im}
\let\Re\relax
\DeclareMathOperator{\Re}{Re}

\let\emptyset\varnothing

\tikzset{->-/.style={decoration={
  markings,
  mark=at position .5 with {\arrow{>}}},postaction={decorate}}}
  \tikzset{-<-/.style={decoration={
  markings,
  mark=at position .5 with {\arrow{<}}},postaction={decorate}}}

\begin{document}
	\title[R\'enyi Negativity for $N$ Disjoint Intervals in $1+1$D Ising CFT]{Replicated Entanglement Negativity for Disjoint Intervals in the Ising Conformal Field Theory}
	\author{Gavin Rockwood}
	\address{Department of Physics and Astronomy, Rutgers University, 136 Frelinghuysen Rd, Piscataway, NJ 08854, US }
	%\keywords{Entanglement Negativity, Entanglment Entropy, Ising, Conformal Field Theory}

	\begin{abstract}
		We calculate the $N$ interval replicated negativity for the Ising conformal field theory using correlation functions of branch point twist fields. For some subset $P\subset N$, this is a calculation of $\Tr(\rho_A^{T_P})^n$ where $n$ is the replica index. This can be reformulated as a calculation of partition functions over superelliptic Riemann surfaces, and for the models in question, this partition function can be expressed in terms of the period matrix of this surface. We detail how to construct the period matrices for these surfaces, giving an analytic expression for $\Tr(\rho_A^{T_P})^n$. The results are expressed such that when $P = \emptyset$, which corresponds to calculating the R\'enyi entropy, the formulas aligns with known results for the $N$ interval R\'enyi entropy.
	\end{abstract}
	\maketitle
	\tableofcontents
	\section{Introduction}
		
		Entanglement entropy has long been known to be an interesting computable quantity in many body systems, it is a measurement of non-local  correlation;. However, it is also a powerful tool to detect critical points in $1+1$D quantum systems. At critical points, systems such as the $1+1$D Ising model are described by conformal field theories (CFTs) and the entanglement entropy of a subsystem $A$, defined as $S_A = \Tr\rho_A\ln\rho_A$ where $\rho_A = \Tr_B\rho$ is the reduced density matrix, goes as $S_A\sim (c/3)\ln \ell$ where $\ell$ is the subsystem  size \citep{calabreseEntanglementEntropyQuantum2004}. This is a useful quantity as it gives the central charge, a fundamental quantity of the CFT. 

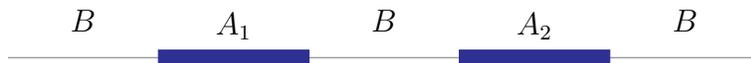
\begin{figure}[H]
			\centering
			\begin{tikzpicture}
				\tikzmath{\lw = 10; \nI = 2; \Dx = \lw/(2*\nI+1); \Dy = 0.2; \sep = 1;}

				\draw[gray] (0,0)--(\lw,0);
				\foreach \i\intervalColor\TFinP in {1/Blue/0,2/Blue/0}
				{
					\filldraw[\intervalColor] ({(2*\i-1)*\Dx},-\Dy/2) rectangle ({(2*\i)*\Dx},\Dy/2);
					\ifthenelse{\TFinP=1}
					{
						\node[above] at ({(2*\i-1+0.5)*\Dx},\Dy/2) {$A_\i\in P$};
					}
					{
						\node[above] at ({(2*\i-1+0.5)*\Dx},\Dy/2) {$A_\i$};
					}
				}
				\node[above] at ({\Dx/2},\Dy) {$B$};
				\node[above] at ({5*\Dx/2},\Dy) {$B$};
				\node[above] at ({9*\Dx/2},\Dy) {$B$};
			\end{tikzpicture}
			\caption{Example with $N=2$.}\label{fig:exampleN=2Configuration}
		\end{figure}

		In general, $S_A$ is hard to calculate analytically. Because of this, we use replicas and calculate the replicated Von-Neumann entropy (or R\'enyi entropy) $S_{A}^{(n)} = \ln\Tr\rho_A^n$. The entanglement entropy is then defined as 
		\begin{equation}
			S_A = \lim_{n\to1}\frac{1}{1-n}S_{A}^{(n)}
		\end{equation}
		The replica trick is useful here because the calculation of $S_A^{(n)}$ is done by calculating correlation functions of branch point twist fields. For certain situations such as those explored in this paper, this calculation reduces to calculating partition functions on certain Riemann surfaces. For free 1+1D CFTs such as the Ising CFT or the free compact boson (FCB), it is well know that their partition functions on genus $g\ge1$ Riemann surfaces are expressible in terms of Riemann Theta functions with dependence on the period matrix $\tau$ of the Riemann surface. This period matrix is relatively straight forward to calculate, meaning that one can write down an analytic (though unfortunately not an algebraic) expression for the partition function \citep{dijkgraafConformalFieldTheories1988,dixonConformalFieldTheory1987}. In past \citep{calabreseEntanglementEntropyConformal2009,calabreseEntanglementEntropyTwo2009,calabreseEntanglementEntropyTwo2011,furukawaMutualInformationBoson2009}, $\Tr\rho^{n}_A$ was calculated for the R\'enyi entropy (for the 1+1D Ising model and FCB) in the case of a two interval bipartition of $\R$, where $\R = A_1\sqcup A_2\sqcup B$ (shown in figure \ref{fig:exampleN=2Configuration}). This gives rise to a bipartite Hilbert space $\H = \H_A \otimes \H_B$, where $\H_A$ is itself a product space $\H_A = \H_{A_1}\otimes\H_{A_2}$ (we will call this the $N = 2$ case). Eventually, this was extended to an $N$-partite Hilbert space $\H_A = \bigotimes_{\gamma=1}^N\H_{A_\gamma}$ in \citep{coserRenyiEntropiesDisjoint2014}.  
		
		Entanglement entropy is not measure of mixed  state entanglement. If one wants to measure the entanglement for mixed states, a useful measure is the logarithmic negativity, $\E = \ln\Tr\abs{\rho^{T_2}}$, where $\rho$ is some mixed state density matrix in a bipartite hilbert space $\H = \H_1\otimes \H_2$ \citep{vidalComputableMeasureEntanglement2002,calabreseEntanglementNegativityQuantum2012}. The partial transpose $T_1$ is the partial transpose with respect to $\H_1$. As a measure of entanglement for mixed states, logarithmic negativity becomes useful when considering systems in thermal mixed states or when degrees of freedom are traced out. In this latter case, we consider a tripartite Hilbert $\H = \H_A\otimes\H_B\otimes\H_C$ with a state 
		\begin{equation}
			\ket{\Psi} = \sum_{i,j,k}\alpha_{i,j,k}\ket{e^i_A}\otimes\ket{e^j_B}\otimes\ket{e^k_C}
		\end{equation}
		where $\{\ket{e_A^i}\}$, $\{\ket{e_B^j}\}$, $\{\ket{e_C^k}\}$ are the bases of $\H_A$, $\H_B$ and $\H_C$ respectively. We now look at the density matrix 
		\begin{equation}
			\begin{aligned}
				\rho &= \ket{\Psi}\bra{\Psi}\\
				&=\sum_{\substack{i,j,k\\i',j',k'}} \alpha_{i,j,k}\alpha^*_{i',j',k'}\ketbra{e^i_A}{e^{i'}_A}\otimes\ketbra{e^j_B}{e^{j'}_B}\otimes\ketbra{e^k_C}{e^{k'}_C}.
			\end{aligned}
		\end{equation}
		The reduced density matrix 
		\begin{equation}\label{eq:ExampleMixedState}
			\begin{aligned}
				\rho_{AB} &= \Tr_C\rho \\
				&= \sum_{\substack{i,j,k\\i',j',k'}}\sum_{m} \alpha_{i,j,k}\alpha^*_{i',j',k'}\ketbra{e^i_A}{e^{i'}_A}\otimes\ketbra{e^j_B}{e^{j'}_B}\times\braOket{e^{m}_C}{\ketbra{e^k_C}{e^{k'}_C}}{e^{m}_C} \\
				&= \sum_{\substack{i,j\\i',j'}} \para{\sum_{k}\alpha_{i,j,k}\alpha^*_{i',j',k}}\ketbra{e^i_A}{e^{i'}_A}\otimes\ketbra{e^j_B}{e^{j'}_B}
			\end{aligned}
		\end{equation}
		is a mixed state density matrix and we want to quantify how entangled this state is. The logarithmic negativity comes from the quantity $\mathcal{N} = \frac{\Tr|\rho_{AB}^{T_2}| - 1}{2}$. This is the negativity and corresponds to the absolute value of the sum of the negative eigenvalues of $\rho_{AB}^{T_2}$. It is well known that if $\rho^{T_2}$ has negative eigenvalues, then the density matrix for that state is non-separable \citep{peresSeparabilityCriterionDensity1996}. This is the PPT (positive partial transpose) criterion and the negativity $\mathcal{N}$ is a measure of how much this criterion is not satisfied. In terms of equation \eqref{eq:ExampleMixedState}, we can ask how far $\rho_{AB}$ is from having the form
		\begin{equation}
			\rho_{AB} = \sum_{k}p^k\rho_A^k\otimes\rho_B^K, \hspace{1cm} p^k\ge0,\;\sum_{k}p^k = 1.
		\end{equation}
		The logarithmic negativity comes into play as it not only is a similar measure but is also an additive quantity under the tensor product of two systems. Meaning that if one has two bipartite systems $\H_\alpha$ and $\H_\beta$ with states $\rho_\alpha$ and $\rho_\beta$, then $\E(\rho_\alpha\otimes\rho_\beta) = \E(\rho_\alpha) + \E(\rho_\beta)$ \citep{vidalComputableMeasureEntanglement2002}\footnote{Note that in some cases, the logarithmic negativity is given as $\log_2\Tr|\rho_A^{T_2}|$ (such as in \citep{vidalComputableMeasureEntanglement2002}). However we will be interested in the natural log instead for consistency with \citep{calabreseEntanglementNegativityQuantum2012,calabreseEntanglementNegativityExtended2013,calabreseEntanglementNegativityCritical2013,nishiokaEntanglementEntropyHolography2018,coserEntanglementNegativityTwo2015}}. Also, we are focusing on $1$D models. This idea of negativity has also been studied in $2$D \cite{eislerEntanglementNegativityTwodimensional2016,denobiliEntanglementNegativityTwo2016}.
		
		In this paper we are interested in the logarithmic negativity of the reduced density matrix $\rho_A$ for the spatial partitions such as the one shown in figure \ref{fig:exampleN=2Configuration}. For this $N=2$ case above, we would want to look at quantities such as $\ln\Tr\abs{\rho_A^{T_{A_2}}}$, which is the negativity between the two intervals of $A$. Unfortunately, like the entanglement entropy, this is not easy to calculate analytically. Instead, we will calculate the replicated logarithmic negativity (what we will call the R\'enyi negativity) $\E^{(n)} = \ln\Tr(\rho_A^{T_{A_2}})^n$. From the R\'enyi negativity, the logarithmic negativity is given by 
		\begin{equation}
			\E = \lim_{n_e\to 1}\E^{(n_e)}.
		\end{equation}
		Note that the analytic continuation is different for even and odd $n$. Because we are interested in the trace norm, the analytic continuation that we will be interested in is the even one as $\Tr\rho_A^{n_e} = \sum\lambda_i^{n_e} = \sum(\lambda_i^2)^{n_e/2} = \sum|\lambda_i|^{n_e}$; the analytic continuation of $\Tr\rho_A^{n_e}$ gives the trace norm in the limit $n_e\to1$. This is a quantity that has been studied in depth in \citep{calabreseEntanglementNegativityQuantum2012,calabreseEntanglementNegativityExtended2013,calabreseEntanglementNegativityCritical2013,nishiokaEntanglementEntropyHolography2018,coserEntanglementNegativityTwo2015,coserPartialTransposeTwo2015} and numerically in \cite{denobiliEntanglementEntropyNegativity2015}. The main goal of this paper is to extend this $N=2$ R\'enyi negativity calculation to the case where $A = \bigsqcup_\gamma^N A_\gamma$ is composed of $N$ disjoint intervals and $\H_A = \bigotimes_\gamma^N\H_{A_\gamma}$ where the choice of which intervals we partially transposing is an arbitrary subset $P$ of the $A_\gamma$s. This is a calculation that again boils down to an expectation value of twist fields which can be expressed in terms of partition function on genus $g = (N-1)(n-1)$ Riemann surfaces, and more explicitly, as a function of the period matrix of the Riemann surface. This is a extending the results of \citep{coserRenyiEntropiesDisjoint2014} to R\'enyi negativity.  

		The paper is organized as follows: In section \ref{sec:SetUp}, we will outline what goes into the calculation of $\Tr\rho_A^{T_P}$ and the notation we will be using. Section \ref{sec:MathStuff} will serve as a quick review of Riemann surfaces and period matrices. We will then replicate the results calculation of $\Tr(\rho_A^{T_2})^n$ for two intervals from \citep{calabreseEntanglementNegativityQuantum2012} by explicitly constructing the period matrix instead of taking advantage of the ``nice'' properties present in this system (which we will discuss). This will naturally lead into the calculation of $\Tr(\rho_A^{T_P})^n$ for general $N$, and $P$. Finally we will explore numerically the limits as we shrink and separate intervals and we will find that in certain cases, quantities like $R_n = \Tr(\rho_A^{T_P})^n/\Tr\rho_A^n$ are continuous up to removable singularities under changes in $N$. This final topic will be the focus of section \ref{sec:ChangeN} while constructions of $\tau$ will be the focus of sections \ref{sec:TauForN} (R\'{e}nyi entropy), \ref{sec:TauForN2} (two interval negativity) and \ref{sec:TauForNP} ($N$ interval negativity). In \ref{sec:NotesTwistFields}, we review some results for twist fields such as the calculation of the central charge and their commutation relations.
		
	\section{Setup}\label{sec:SetUp}
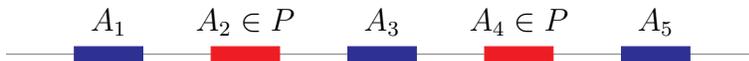
\begin{figure}[H]
			\centering
			\begin{tikzpicture}
				\tikzmath{\lw = 10; \nI = 5; \Dx = \lw/(2*\nI+1); \Dy = 0.2; \sep = 1;}
				\draw[gray] (0,0)--(\lw,0);
				\foreach \i\intervalColor\TFinP in {1/Blue/0,2/Red/1,3/Blue/0,4/Red/1,5/Blue/0}
				{
					\filldraw[\intervalColor] ({(2*\i-1)*\Dx},-\Dy/2) rectangle ({(2*\i)*\Dx},\Dy/2);
					\ifthenelse{\TFinP=1}
					{
						\node[above] at ({(2*\i-1+0.5)*\Dx},\Dy/2) {$A_\i\in P$};
					}
					{
						\node[above] at ({(2*\i-1+0.5)*\Dx},\Dy/2) {$A_\i$};
					}
				}

			\end{tikzpicture}
			\caption{Example with $N=5$ and $P=\{2,4\}$.}\label{fig:exampleN=5Configuration}
		\end{figure}

		In this paper, we will be dealing with a set of $N$ disjoint intervals $A_\gamma$ defined as $A_\gamma = [u_\gamma,v_\gamma]$ with $u_1<v_1<\eli,u_N<v_N$. When discussing logarithmic negativity, we will be partially transposing with respect to a subset $P$ of the $A_\gamma$s (see figure \ref{fig:exampleN=5Configuration} for an $N=5$ with $P = {2,4}$).

		As discussed in \citep{calabreseEntanglementNegativityQuantum2012,calabreseEntanglementNegativityExtended2013}, the R\'{e}nyi entropy for general $N$ is given in terms of expecting value of the branch point twist fields $\T_n(x)$, $\tilde{\T}_n(x)$ (where $n$ is the number of replicas) as\footnote{A quick note on notation: We will generally use Greek letters ($\alpha,\beta,\gamma$) for indexing intervals and Latin indices ($i,j,k$) for indexing sheets.}
		\begin{equation}
		\Tr\rho_A^n = \braket{\prod_{\gamma = 1}^N\T_n(u_\gamma)\tilde{\T}_n(v_\gamma)}_{\hat{\C}}.
		\end{equation}
		For a pair $\T_n(u)\tilde{\T}_n(v)$ (with $u<v$), the resulting path integral expectation value is
		\begin{equation}\label{eq:2TwistFieldExpectation}
			\braket{\T_n(u)\tilde{\T}_n(v)}_{\hat{\C}} = \frac{1}{Z}\int_{C_{u,v}}\prod_{j=1}^n[d\phi_j]e^{-S_{\hat{C}}},
		\end{equation}
		where $C_{u,v}$ is the restriction of the path integral over the $\phi$s to the subspace defined by the locus of the equation $\phi_{j}(x,0^+) = \phi_{j+1}(x,0^{-})$ for $x\in [u,v]$. We will refer to this as the sewing conditions. The label $\hat{\C}$ denotes that this theory is living on the Riemann Sphere $\hat{\C}=\C\cup\{\infty\}$. This label is important as we will eventually move to a multisheeted Riemann surface that we will call $\RNn_{N,n}^P$. For $N$ intervals, equation \eqref{eq:2TwistFieldExpectation} becomes
		\begin{equation}\label{eq:NTwistFieldExpectation}
			\braket{\prod_{\gamma = 1}^N\T_n(u_\gamma)\tilde{\T}_n(v_\gamma)}_{\hat{\C}} = \frac{1}{Z}\int_{C_{u_\gamma,v_\gamma}}\prod_{j=1}^n[d\phi_j]e^{-S_{\hat{C}}},
		\end{equation}
		with the restriction extending to $\phi_{j}(x,0^+) = \phi_{j+1}(x,0^{-})$ for $x\in \bigsqcup_\gamma^N[u_\gamma,v_\gamma]$. For calculating negativity, the intervals in $P$ have their endpoints swapped in the expectation value, meaning $\T_n(u_\gamma)\tilde{\T}_n(v_\gamma)$ becomes $\T_n(v_\gamma)\tilde{\T}_n(u_\gamma)$ for $\gamma \in P$. The path integral restriction then becomes\footnote{Note that the order of twist fields for partially transposed intervals is different than in \citep{calabreseEntanglementNegativityQuantum2012,calabreseEntanglementNegativityCritical2013,calabreseEntanglementNegativityExtended2013}. In these, for a partially transposed interval $\alpha$, one has $\tilde{\T}_n(u_\alpha)\T_n(v_\alpha)$. We will argue that $\brac{\tilde{\T}_n(u_\alpha),\T_n(v_\alpha)}=0$ in \ref{sec:NotesTwistFields}.}
		\begin{equation}\label{eq:NPTwistFieldRestrictions}
			\phi_{j}(x,0^+) = \begin{cases}\phi_{j+1}(x,0^{-}) & x \in \bigsqcup_{\gamma\notin P}[u_\gamma,v_\gamma],\\ \phi_{j-1}(x,0^{-}) & x \in \bigsqcup_{\gamma\in P}[u_\gamma,v_\gamma].\end{cases}
		\end{equation}
		$\Tr(\rho_A^{T_P})^n$ is then given by the expectation value of the twist fields using the new endpoints $\tilde{u}_\gamma,\tilde{v}_\gamma$, defined as $\tilde{u}_\gamma,\tilde{v}_\gamma = u_\gamma,v_\gamma$ for $\gamma \notin P$ and $\tilde{u}_\gamma,\tilde{v}_\gamma = v_\gamma,u_\gamma$ for $\gamma \in P$, form then we can write \eqref{eq:NTwistFieldExpectation} and $\Tr(\rho_A^{T_P})^n$ as
		\begin{equation}\label{eq:NPTwistFieldExpectation}
			\Tr\para{\rho_A^{T_P}}^n = \braket{\prod_{\gamma = 1}^N\T_n(\tilde{u}_\gamma)\tilde{\T}_n(\tilde{v}_\gamma)}_{\hat{\C}} = \frac{1}{Z}\int_{C_{\tilde{u}_\gamma,\tilde{v}_\gamma}}\prod_{j=1}^n[d\phi_j]e^{-S_{\hat{C}}},
		\end{equation}
		where the restrictions $C_{\tilde{u}_\gamma,\tilde{v}_\gamma}$ are the same as those in equation \eqref{eq:NPTwistFieldRestrictions}. The path integral on the RHS of equation \eqref{eq:NPTwistFieldExpectation} can also be written as the path integral of a single field $\phi$ over the genus $g = (N-1)(n-1)$ Riemann surface $\RNn_{N,n}^P$ (see equation \eqref{eq:PathIntegralOverRNnP})\citep{dijkgraafConformalFieldTheories1988,coserRenyiEntropiesDisjoint2014} given in equation \eqref{eq:DefofRNnP} with examples for $N=2,n=3$ shown in figure \ref{fig:N=2ExamplePlots}.
		\begin{equation}\label{eq:PathIntegralOverRNnP}
			\frac{1}{Z}\int_{C_{\tilde{u}_\gamma,\tilde{v}_\gamma}}\prod_{j=1}^n[d\phi_j]e^{-S_{\hat{C}}} = \frac{1}{Z}\int[d\phi]e^{-S_{\RNn_{N,n}^P}} = \frac{Z(\RNn_{N,n}^P)}{Z(\hat{\C})}
		\end{equation}
		\begin{equation}\label{eq:DefofRNnP}
			\RNn_{N,n}^P = \left\{(w,z)\in\hat{\C}\times\hat{\C}\middle|w^n=\prod_{\gamma = 1}^N\para{z-\tilde{u}_\gamma }\para{z-\tilde{v}_\gamma}^{n-1}\right\}
		\end{equation}
		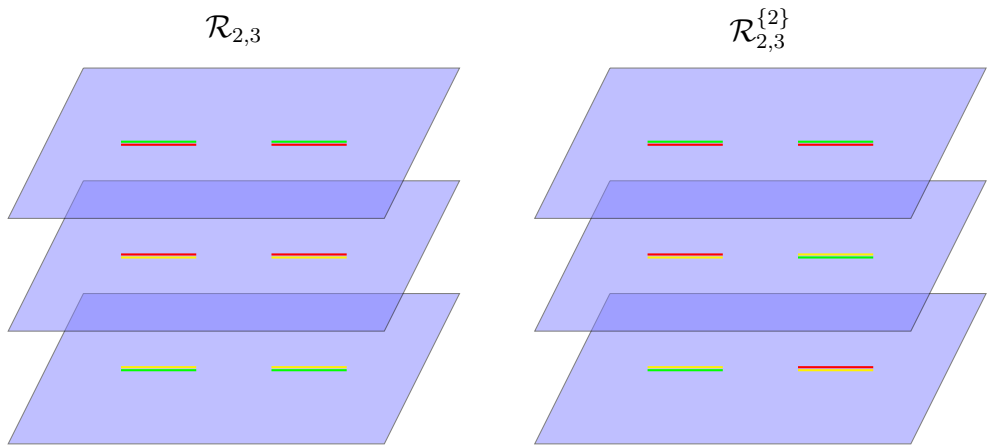
\begin{figure}[h]
			\centering
			\pgfdeclarelayer{1}
			\pgfdeclarelayer{2}
			\pgfdeclarelayer{3}
			\pgfdeclarelayer{labels}
			\pgfsetlayers{3,2,1,labels}
			\begin{tikzpicture}
				\tikzmath{\w= 5; \d = 2; \n = 3; \shift = 1; \tlcx = 0; \tlcy = 0; \dx = 0.75*\d; \sep = 0.02; }
				\foreach \i\j\k in {1/green/red,2/red/yellow,3/yellow/green}
				{ \begin{pgfonlayer}{\i}
						\filldraw[fill = blue!50!white, opacity = 0.5] (\tlcx,{\tlcy -(\i-1)*\dx} ) -- (\tlcx+\w,{\tlcy-(\i-1)*\dx} ) -- (\tlcx+\w-\shift,{\tlcy-\d-(\i-1)*\dx} ) -- (\tlcx-\shift,{\tlcy-\d-(\i-1)*\dx} ) -- (\tlcx,{\tlcy-(\i-1)*\dx} );

						\draw[\j,thick] (\tlcx+0.5,{\tlcy-0.5*\d+\sep-(\i-1)*\dx} ) -- (\tlcx+1.5,{\tlcy-0.5*\d+\sep-(\i-1)*\dx} );
						\draw[\k,thick] (\tlcx+0.5,{\tlcy-0.5*\d-\sep-(\i-1)*\dx} ) -- (\tlcx+1.5,{\tlcy-0.5*\d-\sep-(\i-1)*\dx} );

						\draw[\j,thick] (\tlcx+2.5,{\tlcy-0.5*\d+\sep-(\i-1)*\dx} ) -- (\tlcx+3.5,{\tlcy-0.5*\d+\sep-(\i-1)*\dx} );
						\draw[\k,thick] (\tlcx+2.5,{\tlcy-0.5*\d-\sep-(\i-1)*\dx} ) -- (\tlcx+3.5,{\tlcy-0.5*\d-\sep-(\i-1)*\dx} );
				\end{pgfonlayer}
				}
				\begin{pgfonlayer}{labels}
					\node at (\tlcx+0.5*\w-0.5*\shift, 0.5) {$\RNn_{2,3}$};
				\end{pgfonlayer}

				\tikzmath{\w= 5; \d = 2; \n = 3; \shift = 1; \tlcx = 7; \tlcy = 0; \dx = 0.75*\d; \sep = 0.02;}
				\foreach \i\jL\kL\jR\kR in {1/green/red/green/red,2/red/yellow/yellow/green,3/yellow/green/red/yellow}
				{ \begin{pgfonlayer}{\i}
						\filldraw[fill = blue!50!white, opacity = 0.5] (\tlcx,{\tlcy -(\i-1)*\dx} ) -- (\tlcx+\w,{\tlcy-(\i-1)*\dx} ) -- (\tlcx+\w-\shift,{\tlcy-\d-(\i-1)*\dx} ) -- (\tlcx-\shift,{\tlcy-\d-(\i-1)*\dx} ) -- (\tlcx,{\tlcy-(\i-1)*\dx} );

						\draw[\jL,thick] (\tlcx+0.5,{\tlcy-0.5*\d+\sep-(\i-1)*\dx} ) -- (\tlcx+1.5,{\tlcy-0.5*\d+\sep-(\i-1)*\dx} );
						\draw[\kL,thick] (\tlcx+0.5,{\tlcy-0.5*\d-\sep-(\i-1)*\dx} ) -- (\tlcx+1.5,{\tlcy-0.5*\d-\sep-(\i-1)*\dx} );

						\draw[\jR,thick] (\tlcx+2.5,{\tlcy-0.5*\d+\sep-(\i-1)*\dx} ) -- (\tlcx+3.5,{\tlcy-0.5*\d+\sep-(\i-1)*\dx} );
						\draw[\kR,thick] (\tlcx+2.5,{\tlcy-0.5*\d-\sep-(\i-1)*\dx} ) -- (\tlcx+3.5,{\tlcy-0.5*\d-\sep-(\i-1)*\dx} );
				\end{pgfonlayer}
				}
				\begin{pgfonlayer}{labels}
					\node at (\tlcx+0.5*\w-0.5*\shift, 0.5) {$\RNn_{2,3}^{\{2\}}$};
				\end{pgfonlayer}
			\end{tikzpicture}
			\caption{Plots of $\RNn_{2,3}$ and $\RNn_{2,3}^{\{2\}}$. The colors denote how the branch cuts connect, so for example, moving up through the red side of the first cut on the 1st sheet of $R_{2,n}$ will place you coming out of the $1^\text{st}$ branch on the $2^\text{nd}$ sheet. In general, if there are $N$ branch cuts, moving across the $\gamma$ cut on the $j^\text{th}$ has you coming out of the $\gamma$ cut on the $j\pm1$ sheet (depending on whether or not $\gamma\in P$). $\gamma$ never changes when moving across the branch cuts.}\label{fig:N=2ExamplePlots}
		\end{figure}

		From conformal invariance, we know the expectation value in equation \eqref{eq:NPTwistFieldExpectation} takes on the form\citep{coserRenyiEntropiesDisjoint2014}
		\begin{equation}\label{eq:TrrhoPCFT}
			\Tr(\rho_A^{T_P})^n = \braket{\prod_{\gamma=1}^{N}\T_n(\tilde u_\gamma)\tilde{\T}_n(\tilde v_\gamma)} = c_n^N{\underbrace{\abs{\frac{\prod_{\alpha<\beta}(\tilde u_\beta-\tilde u_\alpha)(\tilde v_\beta-\tilde v_\alpha)}{\prod_{\alpha,\beta}(\tilde v_\beta-\tilde u_\alpha)}}}_{=:\chi(\{[\tilde{u}_\gamma,\tilde{v}_\gamma]\})}}^{2\Delta_{n}}\F^\text{model}(\tau^P_{N,n}),
		\end{equation}
		where $\Delta_n = \tfrac{c}{12}(n - 1/n)$ ($c$ is the central charge, a derivation of this can be found in \ref{sec:NotesTwistFields}) and, $\F$ is both dependent on the model and the surface the theory is defined over. For the Ising CFT ($c=1/2$) and the Free Compact Boson CFT ($c=1$), $\F$ is defined as \eqref{eq:DefofFIsing} and \eqref{eq:DefofFFCB}\cite{dijkgraafConformalFieldTheories1988}:
		\begin{equation}\label{eq:DefofFIsing}
			\F^\text{Ising}(\tau) = \frac{1}{2^g\abs{\rt{0}{\tau}}}\sum\limits_{e,d\in \Z_2^g}\abs{\rtwc{e}{d}{0}{\tau}},
		\end{equation}
		\begin{equation}\label{eq:DefofFFCB}
			\F^{\text{FCB}}(\eta,\tau) = \frac{\rt{0}{G}}{\abs{\rt{0}{\tau}}^2}, \hspace{1cm} G := \begin{pmatrix}i\eta\Im\tau & \Re\tau \\ \Re\tau & i\Im\tau / \eta\end{pmatrix}
		\end{equation}
		where $\eta = \sqrt{2}R$ ($R$ is the compactification radius) and $\tau$ is the period matrix of the Riemann surface the theory is defined over. Therefore, calculating $\tau$ is the main focus of this paper. In addition, equations \eqref{eq:DefofFIsing} and \eqref{eq:DefofFFCB} both involve evaluating Riemann Theta functions over lattices of dimension $g$ and $2g$ respectively. Because of this, we will mainly be making figures for the Ising model as $\F^\text{Ising}(\tau)$ is easier to work with numerically. That said, the general behavior of $\F^{\text{FCB}}(\eta,\tau)$ will be the same.
		
		Also, while we will be mainly applying this to the Ising CFT, if one knows the partition function of some other theory in terms of $\tau$, then one can just follow this construction and apply it to the theory in question\footnote{Such as the Dirac Fermion CFT\cite{dijkgraafConformalFieldTheories1988} or multiple copies of a theory (such as $m$ non-interacting compact bosons).}. Furthermore, the spacetime we start with is $\hat{\C}$. This whole calculation could be extended to R\'enyi entropies or negativities on more complicated surfaces assuming the appropriate basis of holomorphic 1-forms and the canonical homology basis for the resulting Riemann surface are known (for the surfaces in question here, the question of what these are why they are important will discussed in section \ref{sec:MathStuff}).

		Before moving on to calculating $\tau$, it is worth mentioning that a choice of endpoints $\{[u_\gamma,v_\gamma]\}$ does not yield a unique $\tau$ or even $\chi$ due to the conformal symmetry. An easy way to parameterize the problem while removing these redundant degrees of freedom is to instead use the $2N$ harmonic ratios $x_1=h(u_1) = 0$, $x_2 = h(v_1)$, ..., $x_{2\gamma-1} = h(u_\gamma)$, $x_{2\gamma} = h(v_\gamma)$, ..., $x_{2N-3} = h(u_{N-1})$, $x_{2\gamma - 2}$, $x_{2N-1} = h(u_N)= 1$ and $x_{2N} = h(v_N)=\infty$ with
		\begin{equation}\label{eq:DefofHarmonicRatios}
			h(z) = \frac{(z-u_1)(v_N-u_N)}{(u_N - u_1)(v_N - z)}.
		\end{equation}
		Note that while there are $2N$ $x$s, three are fixed, leaving $2N-3$ free parameters with the restriction $0<x_2<\eli,< x_{2N-3}<1$. Using these parameterization is common in the literature \citep{coserRenyiEntropiesDisjoint2014,calabreseEntanglementNegativityExtended2013,coserEntanglementNegativityTwo2015}. However, we will not use this as under partial transposition for $N>2$, there are certain cases where the transformations of the $x$s becomes case dependent such as partially transposing with respect to the first or last interval.

		\subsection{Notes on Riemann Surfaces and Period Matrices}\label{sec:MathStuff}
			In this section we will outline how to obtain the period matrix for a Riemann surface defined by a superelliptic curve $w^n = f(z)$, where $f(z)$ is some polynomial in $z$ of degree $>2$ and with roots $z_\gamma$. For the cases we are interested in, we are studying the $n$ sheeted Riemann surface defined as

			\begin{equation}\label{eq:TheRiemannSurface}
				\RNn_{N,n}^P := \left\{(w,z)\in\hat{\C}\times\hat{\C}\middle|w^n=\prod_{\gamma = 1}^N\para{z-\tilde{u}_\gamma}\para{z-\tilde{v}_\gamma}^{n-1} = \tilde{u}(z)\tilde{v}(z)^{n-1}\right\},
			\end{equation}
			where
			\begin{equation}
				\begin{aligned}
					\tilde{u}(z) &= \prod_{\gamma = 1}^N(z-\tilde{u}_\gamma), &&& \tilde{v}(z) &= \prod_{\gamma = 1}^N(z-\tilde{v}_\gamma).
				\end{aligned}
			\end{equation}
			To calculate the period matrix, we need a basis of holomorphic forms $\omega_{\alpha,j}$ and a canonical homology basis. A basis $\{a_1,\eli,a_{n-1},b_1,\eli,b_{n-1}\}$ of the first homology group $H_1$ is canonical if $a_{\alpha,j}\cdot a_{\beta,k} = b_{\alpha,j}\cdot b_{\beta,k} = 0$ and $a_{\alpha,j}\cdot b_{\beta,k} = - b_{\alpha,j}\cdot a_{\beta,k} = \delta_{\alpha,\beta}\delta_{j,k}$ where the ``$\cdot$'' operation over $H_1$ is the intersection number. The indices $\alpha,\beta$ run over $\{1,\eli,N-1\}$ and $j,k$ run over $\{1,\eli,n-1\}$. Eventually, these two sets of indices will be combined into the indices $r = \beta + (N-1)(k-1)$ and $s=\alpha+(N-1)(j-1)$.

			As in \citep{enolskiSingularCurvesRiemannHilbert2004,coserRenyiEntropiesDisjoint2014}, the holomorphic forms we will use are
			\begin{equation}\label{eq:DefofHolomorphicForms}
				\omega_{\alpha,j} = \frac{z^{\alpha-1}\tilde{v}(z)^{j-1}}{w(z)^{j}}.
			\end{equation}
			If we define the matrices 
			\begin{equation}
				\begin{aligned}
					\A_{rs}=\A_{\beta,k}^{\alpha,j} &= \omega_{\beta,k}(a_{\alpha,j})\\
					\B_{rs}=\B_{\beta,k}^{\alpha,j} &= \omega_{\beta,k}(b_{\alpha,j})
				\end{aligned}
			\end{equation}
			then the period matrix $\tau$ is defined as $\tau := \A^{-1}\B$. $\tau$ will be a symmetric matrix with $\Im\tau$ being positive definite.

			Figuring out what $\omega_{\alpha,j}$ is from some choice of $A_\gamma$s and $P$ is straight forward. The non-trivial component of finding $\tau$ is picking a canonical homology basis. Thus writing down prescription to construct such a basis and expressing $\A$ and $\B$ in terms of that basis is the main bulk of this paper.

			Finally, it is important to note that there is a $\Z_n$ symmetry on $\RNn_{N,n}^P$ generated by the cyclical automorphism $T:w\mapsto e^{2\pi i /n}w$. The interpretation of this automorphism is that it permutes the sheets, or in other words, if one has a curve $\mathcal{C}$ on the $j^\text{th}$ sheet of $\RNn_{N,n}^P$, then $T\mathcal{C}$ is the same curve shifted to the $j+1$ sheet. If we then look at $\omega_{\alpha,j}(T\mathcal{C})$, this is equal to $(T\omega_{\alpha,j})(\mathcal{C})$ and by looking at the definition of $\omega_{\alpha,j}$ in equation \eqref{eq:DefofHolomorphicForms}, we see that $T\omega_{\alpha,j} = e^{-2\pi i j/n}\omega_{\alpha,j}$. Thus $\omega_{\alpha,j}(T\mathcal{C}) = e^{-2\pi j/n}\omega_{\alpha,j}(\mathcal{C})$. This structure will be useful later to simplify the calculations of $\A$ and $\B$.
	
		\subsection{Review of $\tau$ for Entanglement Entropy}\label{sec:TauForN}
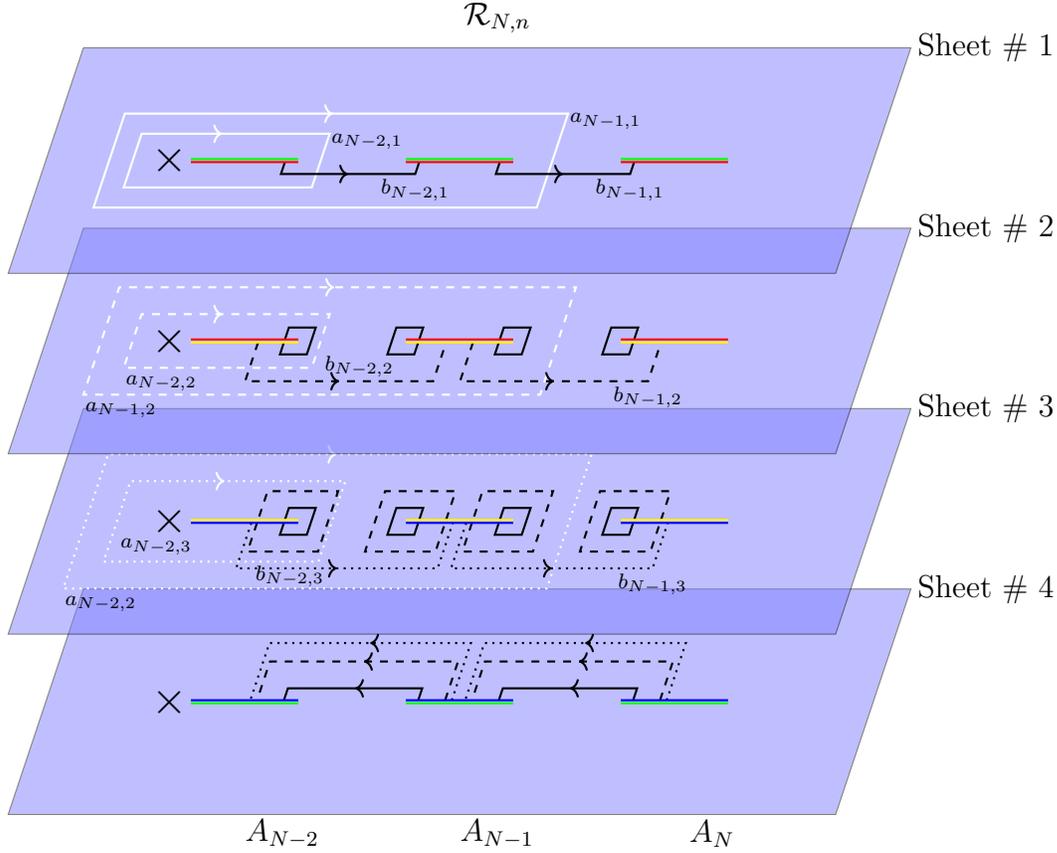
\begin{figure}[h]
				\centering
				\pgfdeclarelayer{1}
				\pgfdeclarelayer{2}
				\pgfdeclarelayer{3}
				\pgfdeclarelayer{4}
				\pgfdeclarelayer{labels}
				\pgfsetlayers{4,3,2,1,labels}
				\begin{tikzpicture}

					\tikzmath{\w= 10; \d = 3; \n = 4; \shift = 1; \tlcx = 0; \tlcy = 0; \Dx = 0.8*\d; \sep = 0.02; \lw = \w/(2*3+1); \ls = 0.25*\lw; \slope = \shift/\d; \pding= 0.05*\w;}
					\foreach \i\jL\kL\jR\kR in {1/green/red/green/red,2/red/yellow/red/yellow,3/yellow/blue/yellow/blue, 4/blue/green/blue/green}
					{ \begin{pgfonlayer}{\i}
							\filldraw[fill = blue!50!white, opacity = 0.5] (\tlcx-\pding,{\tlcy -(\i-1)*\Dx}) -- (\tlcx+\w+\pding,{\tlcy-(\i-1)*\Dx} ) -- (\tlcx+\w-\shift+\pding,{\tlcy-\d-(\i-1)*\Dx} ) -- (\tlcx-\shift-\pding,{\tlcy-\d-(\i-1)*\Dx} ) -- (\tlcx-\pding,{\tlcy-(\i-1)*\Dx} );
						\end{pgfonlayer}

						\begin{pgfonlayer}{labels}
							\node at (\tlcx+1.15*\w, {-(\i-1)*\Dx}) {$\text{Sheet \# }\i$};
						\end{pgfonlayer}

						}

					\begin{pgfonlayer}{labels}
							\node at (5,0.4) {$\RNn_{N,n}$};
					\end{pgfonlayer}
					\begin{pgfonlayer}{labels}
						\node at (1.5*\lw,-3*\Dx-\d-0.25) {$A_{N-2}$};
						\node at (3.5*\lw,-3*\Dx-\d-0.25) {$A_{N-1}$};
						\node at (5.5*\lw,-3*\Dx-\d-0.25) {$A_{N}$};
					\end{pgfonlayer}
						% Alpha 1 basis
					% a11
					\tikzmath{
						\loopsize = 1*\ls;
						\ay = -\loopsize-\d/2+\tlcy;
						\ax = 1.5*\lw+\tlcx-0.5*\shift-\loopsize/2;
						\by = \ay;
						\bx = \ax-0.75*\lw - \loopsize;
						\cy = \by + 2*\loopsize;
						\cx = \bx+(\cy-\by)*\slope;
						\dy = \cy;
						\dx = \cx+1.25*\lw+2*\loopsize;
						\ey = \dy - 2*\loopsize;
						\ex = \dx+(\ey-\dy)*\slope;
						\fy = \ey;
						\fx = \ex-0.5*\lw-\loopsize;
					}
					\begin{pgfonlayer}{1}
						\draw[white, thick,->-] (\ax,\ay)--(\bx,\by)--(\cx,\cy)--(\dx,\dy)--(\ex,\ey)--(\fx,\fy);
					\end{pgfonlayer}
					\begin{pgfonlayer}{labels}
						\node at (\dx+0.5,\dy-0.1) {\scriptsize$a_{N-2,1}$};
					\end{pgfonlayer}

					% a12
					\tikzmath{
						\loopsize = 1*\ls;
						\ay = -\loopsize-\d/2+\tlcy-\Dx;
						\ax = 1.5*\lw+\tlcx-0.5*\shift-\loopsize/2;
						\by = \ay;
						\bx = \ax-0.75*\lw - \loopsize;
						\cy = \by + 2*\loopsize;
						\cx = \bx+(\cy-\by)*\slope;
						\dy = \cy;
						\dx = \cx+1.25*\lw+2*\loopsize;
						\ey = \dy - 2*\loopsize;
						\ex = \dx+(\ey-\dy)*\slope;
						\fy = \ey;
						\fx = \ex-0.5*\lw-\loopsize;
					}
					\begin{pgfonlayer}{2}
						\draw[white, thick, dashed,->-] (\ax,\ay)--(\bx,\by)--(\cx,\cy)--(\dx,\dy)--(\ex,\ey)--(\fx,\fy);
					\end{pgfonlayer}
					\begin{pgfonlayer}{labels}
						\node at (\bx+0.5,\by-0.2) {\scriptsize$a_{N-2,2}$};
					\end{pgfonlayer}

					% a13
					\tikzmath{
						\loopsize = 1.5*\ls;
						\ay = -\loopsize-\d/2+\tlcy-2*\Dx;
						\ax = 1.5*\lw+\tlcx-0.5*\shift-\loopsize/2;
						\by = \ay;
						\bx = \ax-0.75*\lw - \loopsize;
						\cy = \by + 2*\loopsize;
						\cx = \bx+(\cy-\by)*\slope;
						\dy = \cy;
						\dx = \cx+1.25*\lw+2*\loopsize;
						\ey = \dy - 2*\loopsize;
						\ex = \dx+(\ey-\dy)*\slope;
						\fy = \ey;
						\fx = \ex-0.5*\lw-\loopsize;
					}
					\begin{pgfonlayer}{3}
						\draw[white, thick, dotted,->-] (\ax,\ay)--(\bx,\by)--(\cx,\cy)--(\dx,\dy)--(\ex,\ey)--(\fx,\fy);
					\end{pgfonlayer}
					\begin{pgfonlayer}{labels}
						\node at (\bx+0.7,\by+0.2) {\scriptsize$a_{N-2,3}$};
					\end{pgfonlayer}
				% The B Loops
					% b11
					\tikzmath{
						\loopsize = 0.5*\ls;
						\ay = -\sep-\d/2+\tlcy;
						\ax = 2*\lw-\loopsize-\shift/2+\tlcx;
						\by = \ay -1*\loopsize +\sep;
						\bx = \ax+(\by-\ay)*\slope;
						\cy = \by ;
						\cx = \bx + 2*\loopsize+\lw;
						\dy = \cy+\loopsize-\sep;
						\dx = \cx+(\dy-\cy)*\slope;
						\ey = \dy + 2*\sep;
						\ex = \dx;
						\fy = \ey - \sep+\loopsize;
						\fx = \ex+(\fy-\ey)*\slope;
						\gy = \fy;
						\gx = \fx-2*\loopsize;
						\hy = \gy - 2*\loopsize;
						\hx = \gx+(\hy-\gy)*\slope;
						\iy = \hy;
						\ix = \hx + 2*\loopsize;
						\jy = \iy + \loopsize-\sep;
						\jx = \ix + (\jy - \iy)*\slope;
						\ky = \jy + 2*\sep;
						\kx = \jx;
						\ly = \ky + \loopsize-\sep;
						\lx = \kx + (\ly - \ky)*\slope;
						\my = \ly;
						\mx = \lx - 2*\loopsize - \lw;
						\ny = \my - \loopsize + \sep;
						\nx = \mx + (\ny - \my)*\slope;
						\oy = \ny - 2*\sep;
						\ox = \nx;
						\ppy = \oy - \loopsize + \sep;
						\ppx = \ox + (\ppy-\oy)*\slope;
						\qy = \ppy;
						\qx = \ppx+2*\loopsize;
						\ry = \qy + 2*\loopsize;
						\rx = \qx + (\ry-\qy)*\slope;
						\sy = \ry;
						\sx = \rx - 2*\loopsize;
						\ty = \sy - \loopsize+\sep;
						\tx = \sx+(\ty-\sy)*\slope;
					}

					\begin{pgfonlayer}{1}
						\draw[black, thick,->-] (\ax, \ay)--(\bx,\by)--(\cx,\cy)--(\dx,\dy);
					\end{pgfonlayer}
					\begin{pgfonlayer}{2}
						\draw[black, thick] (\ex,\ey-\Dx)--(\fx,\fy-\Dx)--(\gx,\gy-\Dx)--(\hx,\hy-\Dx)--(\ix,\iy-\Dx)--(\jx,\jy-\Dx);
						\draw[black, thick] (\ox, \oy-\Dx)--(\ppx, \ppy-\Dx)--(\qx, \qy-\Dx)--(\rx, \ry-\Dx)--(\sx, \sy-\Dx)--(\tx, \ty-\Dx);
					\end{pgfonlayer}
					\begin{pgfonlayer}{3}
						\draw[black, thick] (\ex,\ey-2*\Dx)--(\fx,\fy-2*\Dx)--(\gx,\gy-2*\Dx)--(\hx,\hy-2*\Dx)--(\ix,\iy-2*\Dx)--(\jx,\jy-2*\Dx);
						\draw[black, thick] (\ox, \oy-2*\Dx)--(\ppx, \ppy-2*\Dx)--(\qx, \qy-2*\Dx)--(\rx, \ry-2*\Dx)--(\sx, \sy-2*\Dx)--(\tx, \ty-2*\Dx);
					\end{pgfonlayer}
					\begin{pgfonlayer}{4}
						\draw[black, thick,->-] (\kx, \ky - 3*\Dx)--(\lx, \ly - 3*\Dx)--(\mx, \my - 3*\Dx)--(\nx, \ny - 3*\Dx);
					\end{pgfonlayer}
					\begin{pgfonlayer}{labels}
						\node at (\cx,\cy-0.21) {\scriptsize$b_{N-2,1}$};
					\end{pgfonlayer}

					% b12
					\tikzmath{
						\loopsize = 1.5*\ls;
						\ay = -\sep-\d/2+\tlcy;
						\ax = 2*\lw-\loopsize-\shift/2+\tlcx;
						\by = \ay -1*\loopsize +\sep;
						\bx = \ax+(\by-\ay)*\slope;
						\cy = \by ;
						\cx = \bx + 2*\loopsize+\lw;
						\dy = \cy+\loopsize-\sep;
						\dx = \cx+(\dy-\cy)*\slope;
						\ey = \dy + 2*\sep;
						\ex = \dx;
						\fy = \ey - \sep+0.75*\loopsize;
						\fx = \ex+(\fy-\ey)*\slope;
						\gy = \fy;
						\gx = \fx-1.75*\loopsize;
						\hy = \gy - 2*0.75*\loopsize;
						\hx = \gx+(\hy-\gy)*\slope;
						\iy = \hy;
						\ix = \hx + 1.75*\loopsize;
						\jy = \iy + 0.75*\loopsize-\sep;
						\jx = \ix + (\jy - \iy)*\slope;
						\ky = \jy + 2*\sep;
						\kx = \jx;
						\ly = \ky + \loopsize-\sep;
						\lx = \kx + (\ly - \ky)*\slope;
						\my = \ly;
						\mx = \lx - 2*\loopsize - \lw;
						\ny = \my - \loopsize + \sep;
						\nx = \mx + (\ny - \my)*\slope;
						\oy = \ny - 2*\sep;
						\ox = \nx;
						\ppy = \oy - 0.75*\loopsize + \sep;
						\ppx = \ox + (\ppy-\oy)*\slope;
						\qy = \ppy;
						\qx = \ppx+1.75*\loopsize;
						\ry = \qy + 2*0.75*\loopsize;
						\rx = \qx + (\ry-\qy)*\slope;
						\sy = \ry;
						\sx = \rx - 1.75*\loopsize;
						\ty = \sy - 0.75*\loopsize+\sep;
						\tx = \sx+(\ty-\sy)*\slope;
					}

					\begin{pgfonlayer}{2}
						\draw[black, thick,dashed,->-] (\ax, \ay-\Dx)--(\bx,\by-\Dx)--(\cx,\cy-\Dx)--(\dx,\dy-\Dx);
					\end{pgfonlayer}
					\begin{pgfonlayer}{3}
						\draw[black, thick, dashed] (\ex,\ey-2*\Dx)--(\fx,\fy-2*\Dx)--(\gx,\gy-2*\Dx)--(\hx,\hy-2*\Dx)--(\ix,\iy-2*\Dx)--(\jx,\jy-2*\Dx);
						\draw[black, thick, dashed] (\ox, \oy-2*\Dx)--(\ppx, \ppy-2*\Dx)--(\qx, \qy-2*\Dx)--(\rx, \ry-2*\Dx)--(\sx, \sy-2*\Dx)--(\tx, \ty-2*\Dx);
					\end{pgfonlayer}
					\begin{pgfonlayer}{4}
						\draw[black, thick,dashed,->-] (\kx, \ky - 3*\Dx)--(\lx, \ly - 3*\Dx)--(\mx, \my - 3*\Dx)--(\nx, \ny - 3*\Dx);
					\end{pgfonlayer}
					\begin{pgfonlayer}{labels}
						\node at (\bx+1.5*\loopsize+\lw/2,\by+0.2-\Dx) {\scriptsize$b_{N-2,2}$};
					\end{pgfonlayer}

				% b13
					\tikzmath{
						\loopsize = 1.75*\ls;
						\ay = -\sep-\d/2+\tlcy;
						\ax = 2*\lw-\loopsize-\shift/2+\tlcx;
						\by = \ay -1*\loopsize +\sep;
						\bx = \ax+(\by-\ay)*\slope;
						\cy = \by ;
						\cx = \bx + 2*\loopsize+\lw;
						\dy = \cy+\loopsize-\sep;
						\dx = \cx+(\dy-\cy)*\slope;
						\ey = \dy + 2*\sep;
						\ex = \dx;
						\fy = \ey - \sep+\loopsize;
						\fx = \ex+(\fy-\ey)*\slope;
						\gy = \fy;
						\gx = \fx-2*\loopsize;
						\hy = \gy - 2*\loopsize;
						\hx = \gx+(\hy-\gy)*\slope;
						\iy = \hy;
						\ix = \hx + 2*\loopsize;
						\jy = \iy + \loopsize-\sep;
						\jx = \ix + (\jy - \iy)*\slope;
						\ky = \jy + 2*\sep;
						\kx = \jx;
						\ly = \ky + 1.25*\loopsize-\sep;
						\lx = \kx + (\ly - \ky)*\slope;
						\my = \ly;
						\mx = \lx - 2*\loopsize - \lw;
						\ny = \my - 1.25*\loopsize + \sep;
						\nx = \mx + (\ny - \my)*\slope;
						\oy = \ny - 2*\sep;
						\ox = \nx;
						\ppy = \oy - \loopsize + \sep;
						\ppx = \ox + (\ppy-\oy)*\slope;
						\qy = \ppy;
						\qx = \ppx+2*\loopsize;
						\ry = \qy + 2*\loopsize;
						\rx = \qx + (\ry-\qy)*\slope;
						\sy = \ry;
						\sx = \rx - 2*\loopsize;
						\ty = \sy - \loopsize+\sep;
						\tx = \sx+(\ty-\sy)*\slope;
					}

					\begin{pgfonlayer}{3}
						\draw[black, thick,dotted,->-] (\ax, \ay-2*\Dx)--(\bx,\by-2*\Dx)--(\cx,\cy-2*\Dx)--(\dx,\dy-2*\Dx);
					\end{pgfonlayer}
					\begin{pgfonlayer}{4}
						\draw[black, thick,dotted,->-] (\kx, \ky - 3*\Dx)--(\lx, \ly - 3*\Dx)--(\mx, \my - 3*\Dx)--(\nx, \ny - 3*\Dx);
					\end{pgfonlayer}
					\begin{pgfonlayer}{labels}
						\node at (\bx+\lw/2,\by-0.1-2*\Dx) {\scriptsize$b_{N-2,3}$};
					\end{pgfonlayer}

					%%%%%%%%
					% Alpha 2 basis
					% a21
					\tikzmath{
						\loopsize = 1.75*\ls;
						\ay = -\loopsize-\d/2+\tlcy;
						\ax = 2.5*\lw+\tlcx-0.5*\shift-\loopsize/2;
						\by = \ay;
						\bx = \ax-1.75*\lw - \loopsize;
						\cy = \by + 2*\loopsize;
						\cx = \bx+(\cy-\by)*\slope;
						\dy = \cy;
						\dx = \cx+3.25*\lw+2*\loopsize;
						\ey = \dy - 2*\loopsize;
						\ex = \dx+(\ey-\dy)*\slope;
						\fy = \ey;
						\fx = \ex-1.5*\lw-\loopsize;
					}
					\begin{pgfonlayer}{1}
						\draw[white, thick,->-] (\ax,\ay)--(\bx,\by)--(\cx,\cy)--(\dx,\dy)--(\ex,\ey)--(\fx,\fy);
					\end{pgfonlayer}
					\begin{pgfonlayer}{labels}
						\node at (\dx+0.5,\dy-0.1) {\scriptsize$a_{N-1,1}$};
					\end{pgfonlayer}

					% a22
					\tikzmath{
						\loopsize = 2*\ls;
						\ay = -\loopsize-\d/2+\tlcy-\Dx;
						\ax = 2.5*\lw+\tlcx-0.5*\shift-\loopsize/2;
						\by = \ay;
						\bx = \ax-1.75*\lw - \loopsize;
						\cy = \by + 2*\loopsize;
						\cx = \bx+(\cy-\by)*\slope;
						\dy = \cy;
						\dx = \cx+3.25*\lw+2*\loopsize;
						\ey = \dy - 2*\loopsize;
						\ex = \dx+(\ey-\dy)*\slope;
						\fy = \ey;
						\fx = \ex-1.5*\lw-\loopsize;
					}
					\begin{pgfonlayer}{2}
						\draw[white, thick, dashed,->-] (\ax,\ay)--(\bx,\by)--(\cx,\cy)--(\dx,\dy)--(\ex,\ey)--(\fx,\fy);
					\end{pgfonlayer}
					\begin{pgfonlayer}{labels}
						\node at (\bx+0.5,\by-0.2) {\scriptsize$a_{N-1,2}$};
					\end{pgfonlayer}

					% a23
					\tikzmath{
						\loopsize = 2.5*\ls;
						\ay = -\loopsize-\d/2+\tlcy-2*\Dx;
						\ax = 2.5*\lw+\tlcx-0.5*\shift-\loopsize/2;
						\by = \ay;
						\bx = \ax-1.75*\lw - \loopsize;
						\cy = \by + 2*\loopsize;
						\cx = \bx+(\cy-\by)*\slope;
						\dy = \cy;
						\dx = \cx+3.25*\lw+2*\loopsize;
						\ey = \dy - 2*\loopsize;
						\ex = \dx+(\ey-\dy)*\slope;
						\fy = \ey;
						\fx = \ex-1.5*\lw-\loopsize;
					}
					\begin{pgfonlayer}{3}
						\draw[white, thick, dotted,->-] (\ax,\ay)--(\bx,\by)--(\cx,\cy)--(\dx,\dy)--(\ex,\ey)--(\fx,\fy);
					\end{pgfonlayer}
					\begin{pgfonlayer}{labels}
						\node at (\bx+0.5,\by-0.2) {\scriptsize$a_{N-2,2}$};
					\end{pgfonlayer}
				% The B 2 Loops
					% b21
					\tikzmath{
						\loopsize = 0.5*\ls;
						\ay = -\sep-\d/2+\tlcy;
						\ax = 4*\lw-\loopsize-\shift/2+\tlcx;
						\by = \ay -1*\loopsize +\sep;
						\bx = \ax+(\by-\ay)*\slope;
						\cy = \by ;
						\cx = \bx + 2*\loopsize+\lw;
						\dy = \cy+\loopsize-\sep;
						\dx = \cx+(\dy-\cy)*\slope;
						\ey = \dy + 2*\sep;
						\ex = \dx;
						\fy = \ey - \sep+\loopsize;
						\fx = \ex+(\fy-\ey)*\slope;
						\gy = \fy;
						\gx = \fx-2*\loopsize;
						\hy = \gy - 2*\loopsize;
						\hx = \gx+(\hy-\gy)*\slope;
						\iy = \hy;
						\ix = \hx + 2*\loopsize;
						\jy = \iy + \loopsize-\sep;
						\jx = \ix + (\jy - \iy)*\slope;
						\ky = \jy + 2*\sep;
						\kx = \jx;
						\ly = \ky + \loopsize-\sep;
						\lx = \kx + (\ly - \ky)*\slope;
						\my = \ly;
						\mx = \lx - 2*\loopsize - \lw;
						\ny = \my - \loopsize + \sep;
						\nx = \mx + (\ny - \my)*\slope;
						\oy = \ny - 2*\sep;
						\ox = \nx;
						\ppy = \oy - \loopsize + \sep;
						\ppx = \ox + (\ppy-\oy)*\slope;
						\qy = \ppy;
						\qx = \ppx+2*\loopsize;
						\ry = \qy + 2*\loopsize;
						\rx = \qx + (\ry-\qy)*\slope;
						\sy = \ry;
						\sx = \rx - 2*\loopsize;
						\ty = \sy - \loopsize+\sep;
						\tx = \sx+(\ty-\sy)*\slope;
					}

					\begin{pgfonlayer}{1}
						\draw[black, thick,->-] (\ax, \ay)--(\bx,\by)--(\cx,\cy)--(\dx,\dy);
					\end{pgfonlayer}
					\begin{pgfonlayer}{2}
						\draw[black, thick] (\ex,\ey-\Dx)--(\fx,\fy-\Dx)--(\gx,\gy-\Dx)--(\hx,\hy-\Dx)--(\ix,\iy-\Dx)--(\jx,\jy-\Dx);
						\draw[black, thick] (\ox, \oy-\Dx)--(\ppx, \ppy-\Dx)--(\qx, \qy-\Dx)--(\rx, \ry-\Dx)--(\sx, \sy-\Dx)--(\tx, \ty-\Dx);
					\end{pgfonlayer}
					\begin{pgfonlayer}{3}
						\draw[black, thick] (\ex,\ey-2*\Dx)--(\fx,\fy-2*\Dx)--(\gx,\gy-2*\Dx)--(\hx,\hy-2*\Dx)--(\ix,\iy-2*\Dx)--(\jx,\jy-2*\Dx);
						\draw[black, thick] (\ox, \oy-2*\Dx)--(\ppx, \ppy-2*\Dx)--(\qx, \qy-2*\Dx)--(\rx, \ry-2*\Dx)--(\sx, \sy-2*\Dx)--(\tx, \ty-2*\Dx);
					\end{pgfonlayer}
					\begin{pgfonlayer}{4}
						\draw[black, thick,->-] (\kx, \ky - 3*\Dx)--(\lx, \ly - 3*\Dx)--(\mx, \my - 3*\Dx)--(\nx, \ny - 3*\Dx);
					\end{pgfonlayer}
					\begin{pgfonlayer}{labels}
						\node at (\cx,\cy-0.21) {\scriptsize$b_{N-1,1}$};
					\end{pgfonlayer}

					% b22
					\tikzmath{
						\loopsize = 1.5*\ls;
						\ay = -\sep-\d/2+\tlcy;
						\ax = 4*\lw-\loopsize-\shift/2+\tlcx;
						\by = \ay -1*\loopsize +\sep;
						\bx = \ax+(\by-\ay)*\slope;
						\cy = \by ;
						\cx = \bx + 2*\loopsize+\lw;
						\dy = \cy+\loopsize-\sep;
						\dx = \cx+(\dy-\cy)*\slope;
						\ey = \dy + 2*\sep;
						\ex = \dx;
						\fy = \ey - \sep+0.75*\loopsize;
						\fx = \ex+(\fy-\ey)*\slope;
						\gy = \fy;
						\gx = \fx-1.75*\loopsize;
						\hy = \gy - 2*0.75*\loopsize;
						\hx = \gx+(\hy-\gy)*\slope;
						\iy = \hy;
						\ix = \hx + 1.75*\loopsize;
						\jy = \iy + 0.75*\loopsize-\sep;
						\jx = \ix + (\jy - \iy)*\slope;
						\ky = \jy + 2*\sep;
						\kx = \jx;
						\ly = \ky + \loopsize-\sep;
						\lx = \kx + (\ly - \ky)*\slope;
						\my = \ly;
						\mx = \lx - 2*\loopsize - \lw;
						\ny = \my - \loopsize + \sep;
						\nx = \mx + (\ny - \my)*\slope;
						\oy = \ny - 2*\sep;
						\ox = \nx;
						\ppy = \oy - 0.75*\loopsize + \sep;
						\ppx = \ox + (\ppy-\oy)*\slope;
						\qy = \ppy;
						\qx = \ppx+1.75*\loopsize;
						\ry = \qy + 2*0.75*\loopsize;
						\rx = \qx + (\ry-\qy)*\slope;
						\sy = \ry;
						\sx = \rx - 1.75*\loopsize;
						\ty = \sy - 0.75*\loopsize+\sep;
						\tx = \sx+(\ty-\sy)*\slope;
					}

					\begin{pgfonlayer}{2}
						\draw[black, thick,dashed,->-] (\ax, \ay-\Dx)--(\bx,\by-\Dx)--(\cx,\cy-\Dx)--(\dx,\dy-\Dx);
					\end{pgfonlayer}
					\begin{pgfonlayer}{3}
						\draw[black, thick, dashed] (\ex,\ey-2*\Dx)--(\fx,\fy-2*\Dx)--(\gx,\gy-2*\Dx)--(\hx,\hy-2*\Dx)--(\ix,\iy-2*\Dx)--(\jx,\jy-2*\Dx);
						\draw[black, thick, dashed] (\ox, \oy-2*\Dx)--(\ppx, \ppy-2*\Dx)--(\qx, \qy-2*\Dx)--(\rx, \ry-2*\Dx)--(\sx, \sy-2*\Dx)--(\tx, \ty-2*\Dx);
					\end{pgfonlayer}
					\begin{pgfonlayer}{4}
						\draw[black, thick,dashed,->-] (\kx, \ky - 3*\Dx)--(\lx, \ly - 3*\Dx)--(\mx, \my - 3*\Dx)--(\nx, \ny - 3*\Dx);
					\end{pgfonlayer}
					\begin{pgfonlayer}{labels}
						\node at (\cx,\cy-0.21-\Dx) {\scriptsize$b_{N-1,2}$};
					\end{pgfonlayer}

				% b23
					\tikzmath{
						\loopsize = 1.75*\ls;
						\ay = -\sep-\d/2+\tlcy;
						\ax = 4*\lw-\loopsize-\shift/2+\tlcx;
						\by = \ay -1*\loopsize +\sep;
						\bx = \ax+(\by-\ay)*\slope;
						\cy = \by ;
						\cx = \bx + 2*\loopsize+\lw;
						\dy = \cy+\loopsize-\sep;
						\dx = \cx+(\dy-\cy)*\slope;
						\ey = \dy + 2*\sep;
						\ex = \dx;
						\fy = \ey - \sep+\loopsize;
						\fx = \ex+(\fy-\ey)*\slope;
						\gy = \fy;
						\gx = \fx-2*\loopsize;
						\hy = \gy - 2*\loopsize;
						\hx = \gx+(\hy-\gy)*\slope;
						\iy = \hy;
						\ix = \hx + 2*\loopsize;
						\jy = \iy + \loopsize-\sep;
						\jx = \ix + (\jy - \iy)*\slope;
						\ky = \jy + 2*\sep;
						\kx = \jx;
						\ly = \ky + 1.25*\loopsize-\sep;
						\lx = \kx + (\ly - \ky)*\slope;
						\my = \ly;
						\mx = \lx - 2*\loopsize - \lw;
						\ny = \my - 1.25*\loopsize + \sep;
						\nx = \mx + (\ny - \my)*\slope;
						\oy = \ny - 2*\sep;
						\ox = \nx;
						\ppy = \oy - \loopsize + \sep;
						\ppx = \ox + (\ppy-\oy)*\slope;
						\qy = \ppy;
						\qx = \ppx+2*\loopsize;
						\ry = \qy + 2*\loopsize;
						\rx = \qx + (\ry-\qy)*\slope;
						\sy = \ry;
						\sx = \rx - 2*\loopsize;
						\ty = \sy - \loopsize+\sep;
						\tx = \sx+(\ty-\sy)*\slope;
					}

					\begin{pgfonlayer}{3}
						\draw[black, thick,dotted,->-] (\ax, \ay-2*\Dx)--(\bx,\by-2*\Dx)--(\cx,\cy-2*\Dx)--(\dx,\dy-2*\Dx);
					\end{pgfonlayer}
					\begin{pgfonlayer}{4}
						\draw[black, thick,dotted,->-] (\kx, \ky - 3*\Dx)--(\lx, \ly - 3*\Dx)--(\mx, \my - 3*\Dx)--(\nx, \ny - 3*\Dx);
					\end{pgfonlayer}
					\begin{pgfonlayer}{labels}
						\node at (\cx,\cy-0.21-2*\Dx) {\scriptsize$b_{N-1,3}$};
					\end{pgfonlayer}

					\foreach \i\jL\kL\jM\kM\jR\kR in {1/green/red/green/red/green/red,2/red/yellow/red/yellow/red/yellow,3/yellow/blue/yellow/blue/yellow/blue, 4/blue/green/blue/green/blue/green}
					{ \begin{pgfonlayer}{\i}

							\draw[\jL,thick] (\tlcx+\lw - 0.5*\shift,{\tlcy-0.5*\d+\sep-(\i-1)*\Dx} ) -- (\tlcx+2*\lw-0.5*\shift,{\tlcy-0.5*\d+\sep-(\i-1)*\Dx} );
							\draw[\kL,thick] (\tlcx+\lw-0.5*\shift,{\tlcy-0.5*\d-\sep-(\i-1)*\Dx} ) -- (\tlcx+2*\lw-0.5*\shift,{\tlcy-0.5*\d-\sep-(\i-1)*\Dx} );

							\draw[\jM,thick] (\tlcx+3*\lw - 0.5*\shift,{\tlcy-0.5*\d+\sep-(\i-1)*\Dx} ) -- (\tlcx+4*\lw-0.5*\shift,{\tlcy-0.5*\d+\sep-(\i-1)*\Dx} );
							\draw[\kM,thick] (\tlcx+3*\lw-0.5*\shift,{\tlcy-0.5*\d-\sep-(\i-1)*\Dx} ) -- (\tlcx+4*\lw-0.5*\shift,{\tlcy-0.5*\d-\sep-(\i-1)*\Dx} );

							\draw[\jR,thick] (\tlcx+5*\lw - 0.5*\shift,{\tlcy-0.5*\d+\sep-(\i-1)*\Dx} ) -- (\tlcx+6*\lw - 0.5*\shift,{\tlcy-0.5*\d+\sep-(\i-1)*\Dx} );
							\draw[\kR,thick] (\tlcx+5*\lw - 0.5*\shift,{\tlcy-0.5*\d-\sep-(\i-1)*\Dx} ) -- (\tlcx+6*\lw - 0.5*\shift,{\tlcy-0.5*\d-\sep-(\i-1)*\Dx} );
						\end{pgfonlayer}
						\begin{pgfonlayer}{labels}
							\node at (\tlcx-\shift/2+0.8*\lw, {-(\i-1)*\Dx-\d/2}) {\Large $\times$};
						\end{pgfonlayer}
						}

				\end{tikzpicture}
				\caption{An example for how one would construct the surface for $N\ge 3$, $n=4$ and $P=\emptyset$. The $a_{\alpha,j}$ loops are in white and the $b_{\alpha,j}$ loops are in black. To construct higher $N$ surfaces, the first $N-3$ branch cuts would be inserted where the ``$\times$'' is such that the last $2$ sets of homology loops go around them.}\label{fig:GeneralNExampleEmptyP}
			\end{figure}

			In this section we review the calculation of $\Tr\rho_A^n$ done in \citep{coserRenyiEntropiesDisjoint2014} that will lay the groundwork for eventually calculating negativity. For this, we will use the basis demonstrated for the $n = 4$, $N\ge 3$ case in figure \ref{fig:GeneralNExampleEmptyP}. To construct this basis, we start with an auxiliary basis $a^{\text{aux}}_{\alpha,j}$ and $b^{\text{aux}}_{\alpha,j}$. The loops $a^{\text{aux}}_{\alpha,j}$ are simply just loops around the $\alpha$ branch cut on the $j^\text{th}$ sheet (in a similar manner as $a_{N-2,-1}$ in figure \ref{fig:GeneralNExampleEmptyP} if no other branch cuts were inserted). The loops $b^{\text{aux}}_{\alpha,j}$ are loops that move up through the $\alpha+1$ branch cut on the $j^\text{th}$ sheet to the $j+1$ sheet, loop counter clockwise back down through the $\alpha$ branch cut and close (in the same manner the $b_{\alpha,3}$ loops in figure \ref{fig:GeneralNExampleEmptyP}). Then we can define the loops $a_{\alpha,j}$ and $b_{\alpha,j}$ as
			\begin{equation}
				\begin{aligned}
					a_{\alpha,j} &= \sum_{\gamma = 1}^\alpha a_{\alpha,j}^{\text{aux}}, &&& b_{\alpha,j} &= \sum_{k = j}^{n-1}b_{\alpha,j}^{\text{aux}}.
				\end{aligned}
			\end{equation}
			See \citep{coserRenyiEntropiesDisjoint2014,enolskiSingularCurvesRiemannHilbert2004} for more discussions on this basis.

			To construct the matrices $\A_{\beta,k}^{\alpha,j} = \omega_{\beta,k}(a_{\alpha,j})$ and $\B_{\beta,k}^{\alpha,j} = \omega_{\beta,k}(b_{\alpha,j})$, we will introduce some paths $c_{\alpha,j}$ and $d_{\alpha,j}$ where $c_{\alpha,j}$ is a straight line connecting $u_{\alpha}$ to $v_{\alpha}$ on the $j^\text{th}$ sheet and $d_{\alpha,j}$ is a line connecting $v_{\alpha}$ to $u_{\alpha+1}$ on the $j^\text{th}$ sheet. With these, we can write
			\begin{equation}\label{eq:OmegaC,D_forEmptyP}
				\begin{aligned}
					\omega_{\beta,k}(a_{\alpha,j}^{\text{aux}}) &= \omega_{\beta,k}(c_{\alpha,j})-\omega_{\beta,k}(c_{\alpha,j+1}),&&& \omega_{\beta,k}(b_{\alpha,j}^{\text{aux}}) &= \omega_{\beta,k}(d_{\alpha,j})-\omega_{\beta,k}(d_{\alpha,j+1}).
				\end{aligned}
			\end{equation}
			We can now use the $T$ automorphism introduced in section \ref{sec:MathStuff} make the simplifications $\omega_{\beta,j}(c_{\alpha,j}) = \omega_{\beta,k}(T^{j-1}c_{\alpha,1}) = e^{-2\pi i k(j-1)/n}\omega_{\beta,k}(c_{\alpha,1})$ and same of the $d_{\alpha,j}$. This allows us to express the integrals in equation \eqref{eq:OmegaC,D_forEmptyP} as
			\begin{equation}
				\begin{aligned}
					\omega_{\beta,k}(a_{\alpha,j}^{\text{aux}}) &= e^{-2\pi i (j-1)k/n}\para{1-e^{-2\pi i k/n}}\int_{u_\alpha}^{v_{\alpha}}\omega_{\beta,k},\\
					\omega_{\beta,k}(b_{\alpha,j}^{\text{aux}}) &= e^{-2\pi i (j-1)k/n}\para{1-e^{-2\pi i k/n}}\int_{v_\alpha}^{u_{\alpha+1}}\omega_{\beta,k}.
				\end{aligned}
			\end{equation}
			Going back to the $a,b$ basis, we have
			\begin{equation}
				\begin{aligned}
					\A_{\beta,k}^{\alpha,j}=\omega_{\beta,k}(a_{\alpha,j}) &= e^{-2\pi i (j-1)k/n}\para{1-e^{-2\pi i k/n}}\sum_{\gamma = 1}^{\alpha}\int_{u_\gamma}^{v_{\gamma}}\omega_{\beta,k},\\
					\B_{\beta,k}^{\alpha,j}=\omega_{\beta,k}(b_{\alpha,j}) &= \sum_{l = j}^{n-1}e^{-2\pi i (l-1)k/n}\para{1-e^{-2\pi i k/n}}\int_{v_\alpha}^{u_{\alpha+1}}\omega_{\beta,k}\\
					&=\para{e^{-2\pi i k(j-1)/n}-e^{-2\pi i k(n-1)/n}}\int_{v_\alpha}^{u_{\alpha+1}}\omega_{\beta,k}.
				\end{aligned}
			\end{equation}
			As mentioned earlier, this is a result of \citep{coserRenyiEntropiesDisjoint2014}.
	
		\subsection{Calculating $\tau$ for Two Interval Negativity}\label{sec:TauForN2}
\begin{figure}[H]
				\centering
				\pgfdeclarelayer{1}
				\pgfdeclarelayer{2}
				\pgfdeclarelayer{3}
				\pgfdeclarelayer{4}
				\pgfdeclarelayer{labels}
				\pgfsetlayers{4,3,2,1,labels}
				\begin{tikzpicture}
					\tikzmath{\w= 10; \d = 2; \n = 4; \shift = 1; \tlcx = 0; \tlcy = 0; \Dx = 0.8*\d; \sep = 0.02; \lw = \w/(2+3); \ls = 0.25; \slope = 2*\d/\w;}
					\foreach \i\jL\kL\jR\kR in {1/green/red/green/red,2/red/yellow/blue/green,3/yellow/blue/yellow/blue, 4/blue/green/red/yellow}
					{ \begin{pgfonlayer}{\i}
							\filldraw[fill = blue!50!white, opacity = 0.5] (\tlcx,{\tlcy -(\i-1)*\Dx} ) -- (\tlcx+\w,{\tlcy-(\i-1)*\Dx} ) -- (\tlcx+\w-\shift,{\tlcy-\d-(\i-1)*\Dx} ) -- (\tlcx-\shift,{\tlcy-\d-(\i-1)*\Dx} ) -- (\tlcx,{\tlcy-(\i-1)*\Dx} );
						\end{pgfonlayer}
						}

					% THE A BASIS:
						% a_1
						\tikzmath{
							\loopsize = \ls;
							\ay = -\sep-\d/2;
							\ax = \lw+\loopsize-\shift/2;
							\by = \ay + \sep - \loopsize;
							\bx = \ax+(\by-\ay)*\slope;
							\cy = \by;
							\cx = \bx - 2*\loopsize;
							\dy = \cy + 2*\loopsize;
							\dx = \cx+(\dy-\cy)*\slope;
							\ey = \dy;
							\ex = \dx + \lw + 2*\loopsize;
							\fy = \ey - 2*\loopsize;
							\fx = \ex+(\fy-\ey)*\slope;
							\gy = \fy;
							\gx = \fx - 2*\loopsize;
							\hy = \gy + \loopsize - \sep;
							\hx = \gx+(\hy-\gy)*\slope;
							\iy = \hy+2*\sep;
							\ix = \hx;
							\jy = \iy - \sep+\loopsize;
							\jx = \ix+(\jy-\iy)*\slope;
							\ky = \jy - \loopsize;
							\kx = \jx + 2*\loopsize;
							\ly = \ky - \loopsize;
							\lx = \kx - 2*\loopsize+(\ly-\jy)*\slope;
							\my = \ly +\loopsize -\sep;
							\mx = \lx+(\my-\ly)*\slope;
							\ny = \my+2*\sep;
							\nx = \mx;
							\oy = \ny -\sep + \loopsize;
							\ox = \nx+(\oy-\ny)*\slope;
							\ppy = \oy - \loopsize;
							\ppx = \ox+2*\loopsize;
							\qy = \ppy - \loopsize;
							\qx = \ppx - 2*\loopsize+(\qy-\oy)*\slope;
							\ry = \qy;
							\rx = \qx - \lw;
							\sy = \ry + 2*\loopsize;
							\sx = \rx+(\sy-\ry)*\slope;
							\ty = \sy;
							\tx = \sx+\loopsize+(\ty - \ry)*\slope;
							\uy = \ty - \loopsize+\sep;
							\ux = \tx+(\uy-\ty)*\slope;
							\vy = \uy - 2*\sep;
							\vx = \ux;
							\wy = \vy + \sep - \loopsize;
							\wx = \vx+(\wy-\vy)*\slope;
							\xy = \wy;
							\xx = \wx-2*\loopsize;
							\yy = \xy+2*\loopsize;
							\yx = \xx+(\yy-\xy)*\slope;
							\zy = \yy;
							\zx = \yx+2*\loopsize;
							\aay = \zy - \loopsize+\sep;
							\aax = \zx+(\aay-\zy)*\slope;
						}
						\begin{pgfonlayer}{1}
							\draw[white, thick,->-] (\ax,\ay)--(\bx,\by)--(\cx,\cy)--(\dx,\dy)--(\ex,\ey)--(\fx,\fy)--(\gx,\gy)--(\hx,\hy);
						\end{pgfonlayer}
						\begin{pgfonlayer}{2}
							\draw[white, thick] (\ix, \iy-\Dx)-- (\jx,\jy-\Dx) -- (\kx, \ky-\Dx) -- (\lx,\ly-\Dx) -- (\mx, \my-\Dx);
							\draw[white, thick] (\vx, \vy-\Dx) -- (\wx, \wy-\Dx) -- (\xx, \xy - \Dx) -- (\yx, \yy-\Dx) -- (\zx, \zy - \Dx) -- (\aax, \aay - \Dx);
						\end{pgfonlayer}
						\begin{pgfonlayer}{3}
							\draw[white, thick,->-] (\nx, \ny-2*\Dx) -- (\ox, \oy - 2*\Dx)--(\ppx,\ppy-2*\Dx)-- (\qx, \qy-2*\Dx)--(\rx,\ry-2*\Dx)--(\sx,\sy-2*\Dx)--(\tx,\ty-2*\Dx)--(\ux,\uy-2*\Dx);
						\end{pgfonlayer}

						% a_2
						\tikzmath{
							\loopsize = 2*\ls;
							\ay = -\sep-\d/2;
							\ax = \lw+\loopsize-\shift/2;
							\by = \ay + \sep - \loopsize;
							\bx = \ax+(\by-\ay)*\slope;
							\cy = \by;
							\cx = \bx - 2*\loopsize;
							\dy = \cy + 2*\loopsize;
							\dx = \cx+(\dy-\cy)*\slope;
							\ey = \dy;
							\ex = \dx + \lw + 2*\loopsize;
							\fy = \ey - 2*\loopsize;
							\fx = \ex+(\fy-\ey)*\slope;
							\gy = \fy;
							\gx = \fx - 2*\loopsize;
							\hy = \gy + \loopsize - \sep;
							\hx = \gx+(\hy-\gy)*\slope;
							\ny = \hy+2*\sep;
							\nx = \hx;
							\oy = \ny -\sep + \loopsize;
							\ox = \nx+(\oy-\ny)*\slope;
							\ppy = \oy - \loopsize;
							\ppx = \ox+2*\loopsize;
							\qy = \ppy - \loopsize;
							\qx = \ppx - 2*\loopsize+(\qy-\oy)*\slope;
							\ry = \qy;
							\rx = \qx - \lw;
							\sy = \ry + 2*\loopsize;
							\sx = \rx+(\sy-\ry)*\slope;
							\ty = \sy;
							\tx = \sx+2*\loopsize;
							\uy = \ty - \loopsize+\sep;
							\ux = \tx+(\uy-\ty)*\slope;
						}
						\begin{pgfonlayer}{2}
							\draw[white, thick, dashed, ->-] (\ax,\ay-\Dx)--(\bx,\by-\Dx)--(\cx,\cy-\Dx)--(\dx,\dy-\Dx)--(\ex,\ey-\Dx)--(\fx,\fy-\Dx)--(\gx,\gy-\Dx)--(\hx,\hy-\Dx);
						\end{pgfonlayer}
						\begin{pgfonlayer}{3}
							\draw[white, thick, dashed,->-] (\nx, \ny-2*\Dx) -- (\ox, \oy - 2*\Dx)--(\ppx,\ppy-2*\Dx)-- (\qx, \qy-2*\Dx)--(\rx,\ry-2*\Dx)--(\sx,\sy-2*\Dx)--(\tx,\ty-2*\Dx)--(\ux,\uy-2*\Dx);
						\end{pgfonlayer}

						% a_3
						\tikzmath{
							\loopsize = 3*\ls;
							\ay = \loopsize-\d/2;
							\ax = 1.5*\lw-\loopsize*\slope;
							\by = \ay;
							\bx = \ax+1*\loopsize+0.5*\lw+\loopsize*\slope;
							\cy = \by-2*\loopsize;
							\cx = \bx+(\cy-\by)*\slope;
							\dy = \cy;
							\dx = \cx-2*\loopsize-\lw-\loopsize*\slope;
							\ey = \dy + 2*\loopsize;
							\ex = \dx+(\ey-\dy)*\slope;
							\fy = \ey;
							\fx = \ex+\loopsize+0.5*\lw;
						}
						\begin{pgfonlayer}{3}
							\draw[white, thick, dotted,->-] (\ax, \ay-2*\Dx) -- (\bx,\by - 2*\Dx)--(\cx,\cy-2*\Dx)-- (\dx, \dy-2*\Dx)--(\ex,\ey-2*\Dx)--(\fx,\fy-2*\Dx);
						\end{pgfonlayer}
					% The B Loops
						% b1
						\tikzmath{
							\loopsize = 1*\ls;
							\ay = \sep-\d/2;
							\ax = 2*\lw-\loopsize-\shift/2;
							\by = \ay +1*\loopsize -\sep-0.05;
							\bx = \ax+(\by-\ay)*\slope;
							\cy = \by - 2*\loopsize+0.05;
							\cx = \bx + 2*\loopsize+\lw+(\cy-\by)*\slope;
							\dy = \cy+\loopsize-\sep;
							\dx = \cx+(\dy-\cy)*\slope;
							\ey = \dy + \sep;
							\ex = \dx;
							\fy = \ey - \sep+\loopsize;
							\fx = \ex+(\fy-\ey)*\slope;
							\gy = \fy;
							\gx = \fx-2*\loopsize-2*\lw;
							\hy = \gy - 2*\loopsize;
							\hx = \gx+(\hy-\gy)*\slope;
							\iy = \hy;
							\ix = \hx + \lw;
							\jy = \iy+\loopsize-\sep;
							\jx = \ix+(\jy-\iy)*\slope;
						}

						\begin{pgfonlayer}{1}
							\draw[black, thick,->-] (\ax, \ay)--(\bx,\by)--(\cx,\cy)--(\dx,\dy);
						\end{pgfonlayer}
						\begin{pgfonlayer}{4}
							\draw[black, thick,->-] (\ex,\ey-3*\Dx)--(\fx,\fy-3*\Dx)--(\gx,\gy-3*\Dx)--(\hx,\hy-3*\Dx)--(\ix,\iy-3*\Dx)--(\jx,\jy-3*\Dx);
						\end{pgfonlayer}

						% b2
						\tikzmath{
						\loopsize = 2*\ls;
						\ay = \sep-\d/2;
						\ax = 2*\lw-\loopsize-\shift/2;
						\by = \ay +1*\loopsize -\sep-0.05;
						\bx = \ax+(\by-\ay)*\slope;
						\cy = \by - 2*\loopsize+0.05;
						\cx = \bx + 2*\loopsize+\lw+(\cy-\by)*\slope;
						\dy = \cy+\loopsize-\sep;
						\dx = \cx+(\dy-\cy)*\slope;
						\ey = \dy + \sep;
						\ex = \dx;
						\fy = \ey - \sep+\loopsize;
						\fx = \ex+(\fy-\ey)*\slope;
						\gy = \fy;
						\gx = \fx-2*\loopsize-2*\lw;
						\hy = \gy - 2*\loopsize;
						\hx = \gx+(\hy-\gy)*\slope;
						\iy = \hy;
						\ix = \hx + \lw;
						\jy = \iy+\loopsize-\sep;
						\jx = \ix+(\jy-\iy)*\slope;
					}

					\begin{pgfonlayer}{2}
						\draw[black, thick, dashed,->-] (\ax, \ay-\Dx)--(\bx,\by-\Dx)--(\cx,\cy-\Dx)--(\dx,\dy-\Dx);
					\end{pgfonlayer}
					\begin{pgfonlayer}{1}
						\draw[black, thick, dashed,->-] (\ex,\ey)--(\fx,\fy)--(\gx,\gy)--(\hx,\hy)--(\ix,\iy)--(\jx,\jy);
					\end{pgfonlayer}

					% b3
					\tikzmath{
						\loopsize = 3*\ls;
						\ay = \sep-\d/2;
						\ax = 2*\lw-\loopsize-\shift/2;
						\by = \ay +1*\loopsize -\sep-0.05;
						\bx = \ax+(\by-\ay)*\slope;
						\cy = \by - 2*\loopsize+0.05;
						\cx = \bx + 2*\loopsize+\lw+(\cy-\by)*\slope;
						\dy = \cy+\loopsize-\sep;
						\dx = \cx+(\dy-\cy)*\slope;
						\ey = \dy + \sep;
						\ex = \dx;
						\fy = \ey - \sep+\loopsize;
						\fx = \ex+(\fy-\ey)*\slope;
						\gy = \fy;
						\gx = \fx-2*\loopsize-2*\lw;
						\hy = \gy - 2*\loopsize+0.1;
						\hx = \gx+(\hy-\gy)*\slope;
						\iy = \hy;
						\ix = \hx + \lw;
						\jy = \iy+\loopsize-\sep-0.1;
						\jx = \ix+(\jy-\iy)*\slope;
					}

					\begin{pgfonlayer}{3}
						\draw[black, thick, dotted,->-] (\ax, \ay-2*\Dx)--(\bx,\by-2*\Dx)--(\cx,\cy-2*\Dx)--(\dx,\dy-2*\Dx);
					\end{pgfonlayer}
					\begin{pgfonlayer}{2}
						\draw[black, thick, dotted,->-] (\ex,\ey-\Dx)--(\fx,\fy-\Dx)--(\gx,\gy-\Dx)--(\hx,\hy-\Dx)--(\ix,\iy-\Dx)--(\jx,\jy-\Dx);
					\end{pgfonlayer}
					\begin{pgfonlayer}{labels}
						\node at (\tlcx+0.5*\w-0.5*\shift, 0.5) {$\RNn_{2,3}^{\{2\}}$};
						\node[white] at (2*\lw - 0.5*\shift+0.5*\ls, -\d/2-1.75*\ls) {$a_{1,1}$};
						\node[white] at (2*\lw - 0.5*\shift+1.5*\ls, -\d/2-2.75*\ls-\Dx) {$a_{1,2}$};
						\node[white] at (2*\lw - 0.5*\shift+4*\ls, -\d/2-3.25*\ls-2*\Dx) {$a_{1,3}$};

						\node[black] at (2.5*\lw - 0.5*\shift, -\d/2-1.75*\ls) {$b_{1,1}$};
						\node[black] at (2.5*\lw - 0.5*\shift, -\d/2-1.75*\ls-\Dx) {$b_{1,2}$};
						\node[black] at (3*\lw - 0.5*\shift+3.75*\ls, -\d/2-1.75*\ls-2*\Dx) {$b_{1,3}$};
					\end{pgfonlayer}

					\foreach \i\jL\kL\jR\kR in {1/green/red/green/red,2/red/yellow/blue/green,3/yellow/blue/yellow/blue, 4/blue/green/red/yellow}
					{ \begin{pgfonlayer}{\i}

							\draw[\jL,thick] (\tlcx+\lw - 0.5*\shift,{\tlcy-0.5*\d+\sep-(\i-1)*\Dx} ) -- (\tlcx+2*\lw-0.5*\shift,{\tlcy-0.5*\d+\sep-(\i-1)*\Dx} );
							\draw[\kL,thick] (\tlcx+\lw-0.5*\shift,{\tlcy-0.5*\d-\sep-(\i-1)*\Dx} ) -- (\tlcx+2*\lw-0.5*\shift,{\tlcy-0.5*\d-\sep-(\i-1)*\Dx} );

							\draw[\jR,thick] (\tlcx+3*\lw - 0.5*\shift,{\tlcy-0.5*\d+\sep-(\i-1)*\Dx} ) -- (\tlcx+4*\lw - 0.5*\shift,{\tlcy-0.5*\d+\sep-(\i-1)*\Dx} );
							\draw[\kR,thick] (\tlcx+3*\lw - 0.5*\shift,{\tlcy-0.5*\d-\sep-(\i-1)*\Dx} ) -- (\tlcx+4*\lw - 0.5*\shift,{\tlcy-0.5*\d-\sep-(\i-1)*\Dx} );
						\end{pgfonlayer}

						\begin{pgfonlayer}{labels}
							\node at (\tlcx+1.1*\w, {-(\i-1)*\Dx}) {$j=\i$};
						\end{pgfonlayer}
						}
				\end{tikzpicture}
				\caption{The surface $\RNn_{2,3}^{\{2\}}$}\label{fig:N=2n=4Surface}
			\end{figure}
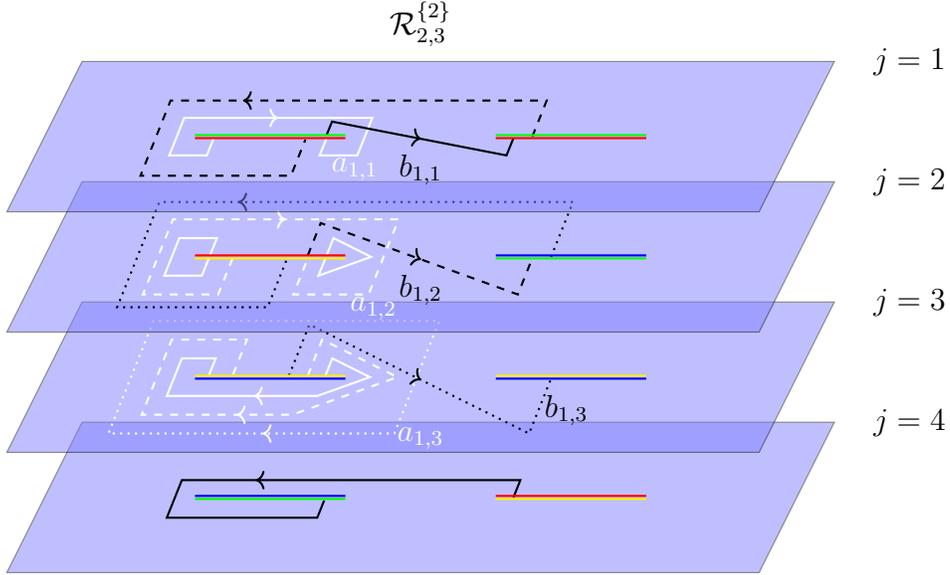

			Now that we have an understanding of the period matrix for entanglement entropy, we are going to focus on the simplest case: $N=2$. We will do this for two reasons. Firstly, the integrals $\omega_{k}(a_j)$ and $\omega_{k}(b_j)$ have analytic solutions for all $n$ (note, because $N=2$, the $\beta,\alpha$ indices are just $1$ and thus we drop them). Secondly, when expressed in terms of the single free harmonic ratio $0<x<1$, there simple relation between the period matrix for entropy and the period matrix for negativity when one partially transposes the second interval\citep{calabreseEntanglementNegativityQuantum2012}. The period for entropy ($\RNn_{2,n}$)is given as
			\begin{equation}\label{eq:N2TauClosed}
				\tau_{kj}(x) = \frac{2i}{n}\sum_{k=1}^{n-1}\sin\para{\frac{\pi k}{n}}\frac{{}_2F_1\para{k/n,1-k/n;1;1-x}}{{}_2F_1\para{k/n,1-k/n;1;x}}\cos\para{\frac{2\pi i k}{n}(k-j)}.
			\end{equation}
			The period matrix for negativity ($\RNn_{2,n}^{\{2\}}$) is then just given $\tau^{\{2\}}(x) = \tau\para{\tfrac{x}{x-1}}$. However, this result does not generalize to arbitrary $N$ and $P$. To do that, we need to adapt the calculations for $\tau(\RNn_{N,n})$ in section \ref{sec:TauForN} to $\RNn_{N,n}^P$. The first step is to construct the $\tau$ for the case described above as it is simple and, as we will see later, provides an easy method of generalization. An example of the basis we will use is given in equation \eqref{fig:N=2n=4Surface} for $n=4$.

\begin{figure}[H]
				\centering
				\includegraphics[scale = 0.6]{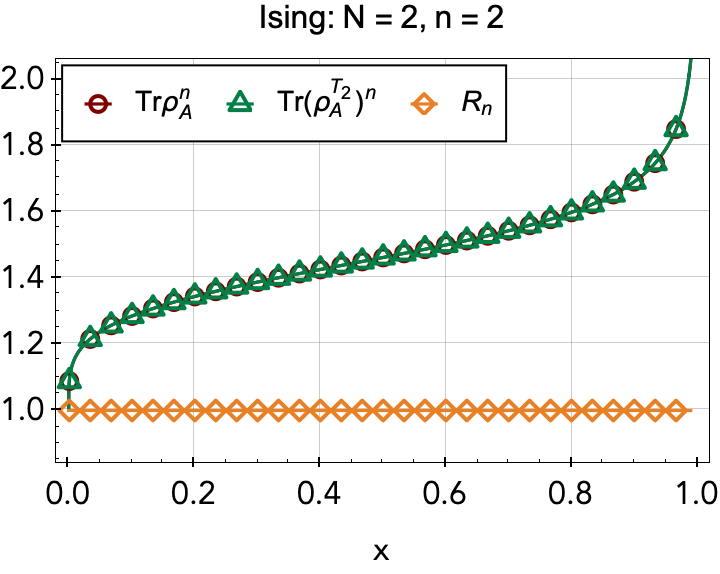}
				\includegraphics[scale = 0.6]{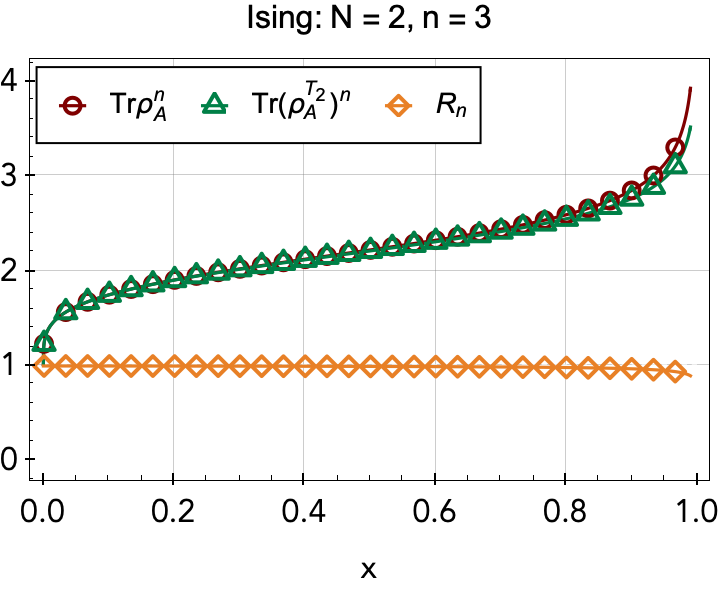}
				\includegraphics[scale = 0.6]{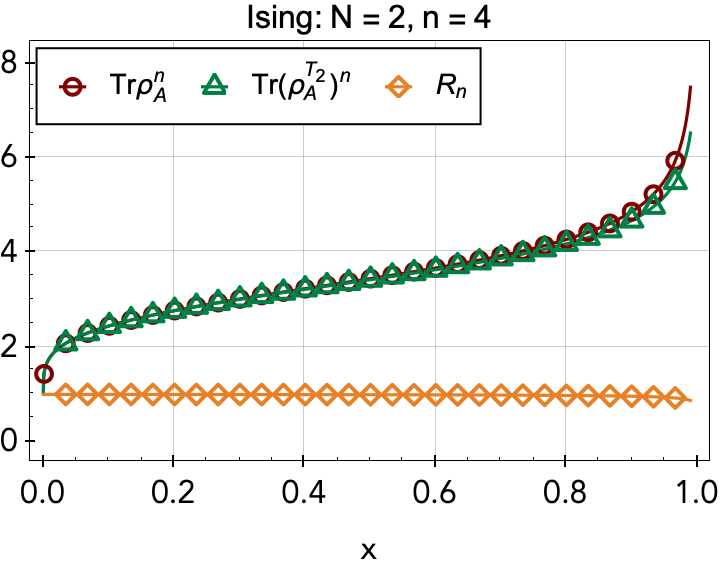}
				\includegraphics[scale=0.6]{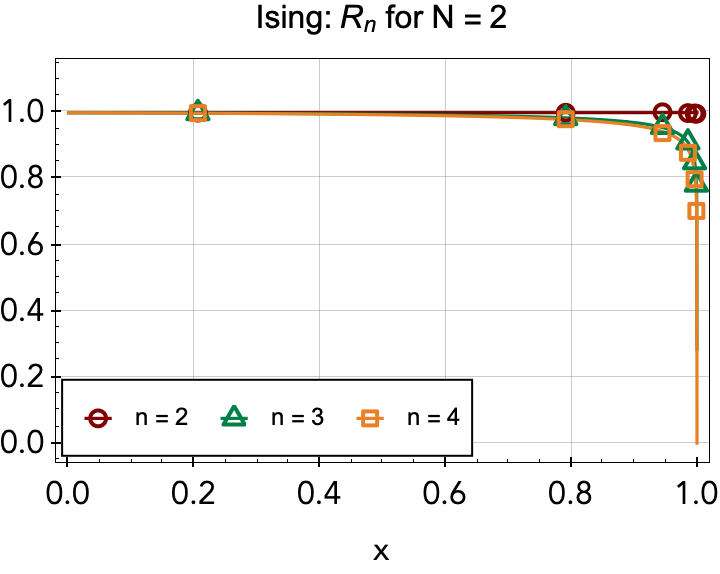}
				\caption{Plots of $\Tr\rho_A^n$, $\Tr(\rho_A^{T_2})^n$ and $R_n = \Tr(\rho_A^{T_2})^n/\Tr(\rho_A)^n$ for the Ising model. Solid lines are the analytic $\tau$ from \eqref{eq:N2TauClosed} and the markers numerical integration of \eqref{eq:N=2AandB}.}\label{fig:IsingN=2Plots}
			\end{figure}

\begin{figure}[H]
				\centering
				\includegraphics[scale = 0.6]{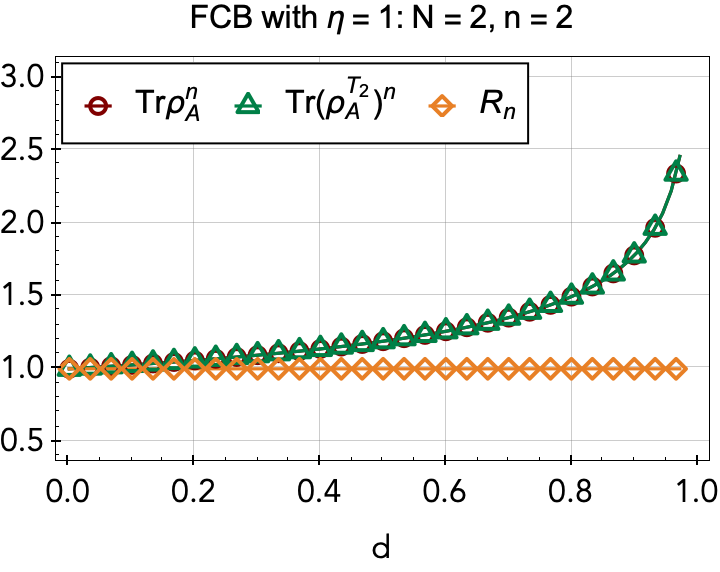}
				\includegraphics[scale = 0.6]{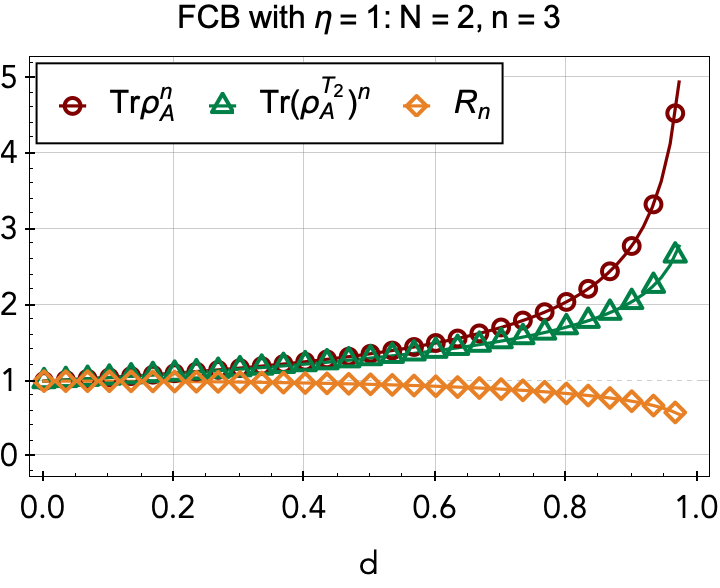}
				\includegraphics[scale=0.6]{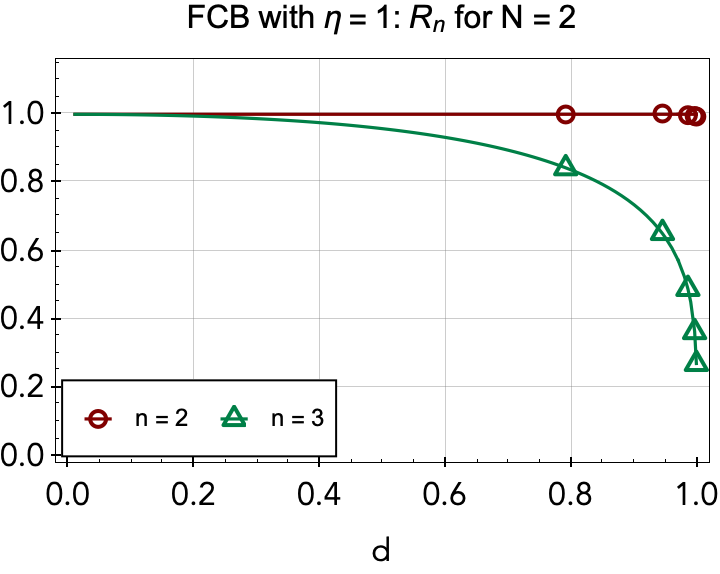}
				\caption{Plots of $\Tr\rho_A^n$, $\Tr(\rho_A^{T_2})^n$ and $R_n = \Tr(\rho_A^{T_2})^n/\Tr(\rho_A)^n$ for the free compact boson with $\eta = 1$. Solid lines are the analytic $\tau$ from \eqref{eq:N2TauClosed} and the markers numerical integration of \eqref{eq:N=2AandB}.}\label{fig:FCBN=2Plots}
			\end{figure}

			The integrals $\omega_{k}(a_j)$ and $\omega_{k}(b_j)$ can again be expressed in terms of $\omega_k(c_j)$ and $\omega_k(d_j)$ respectively to give
			\begin{equation}\label{eq:N=2AandB}
				\begin{aligned}
					\A_{kj} &= \sum_{l=j}^{n-1}\para{\omega_k(c_l)-\omega_k(c_{l+1})} = \para{1-e^{-2\pi i k/n}}\sum\limits_{l=j}^{n-1}\para{e^{-2\pi i  (l-1)/k}}\int_{u_1}^{v_1}\omega_k,\\
					\B_{kj} &=\omega_k(d_j)-\omega_k(d_{j-1}) = \para{e^{-2\pi i (j-1)k/n}-e^{-2\pi i (j-2)k/n}}\int_{v_1}^{u_2}\omega_k.
				\end{aligned}
			\end{equation}
			$\A_{kj}$ can be further simplified using
			\begin{equation}\label{eq:PhaseSumFormula1}
				\sum\limits_{l=j}^{n-1}\para{e^{-2\pi i  (l-1)/k}} = \frac{e^{-2 \pi  i j k/n}-1}{e^{-2 \pi  i k/n} \left(1-e^{-2 \pi  i k/n}\right)}.
			\end{equation}
			Plotting $\Tr\rho_A^n$ and $\Tr\para{\rho_A^{T_2}}^n$ (given in equation \eqref{eq:TrrhoPCFT}) using both the period matrix generated by $\A^{-1}\B$ from equation \eqref{eq:N=2AandB} and the analytic form from \eqref{eq:N2TauClosed} as a function of the free harmonic ratio $x_3 = x$ for the Ising model and free compact boson gives figures \ref{fig:IsingN=2Plots} and \ref{fig:FCBN=2Plots} respectively.

	\section{$\tau$ for General $N$ Interval Entanglement Negativity}\label{sec:TauForNP}

\begin{figure}[H]
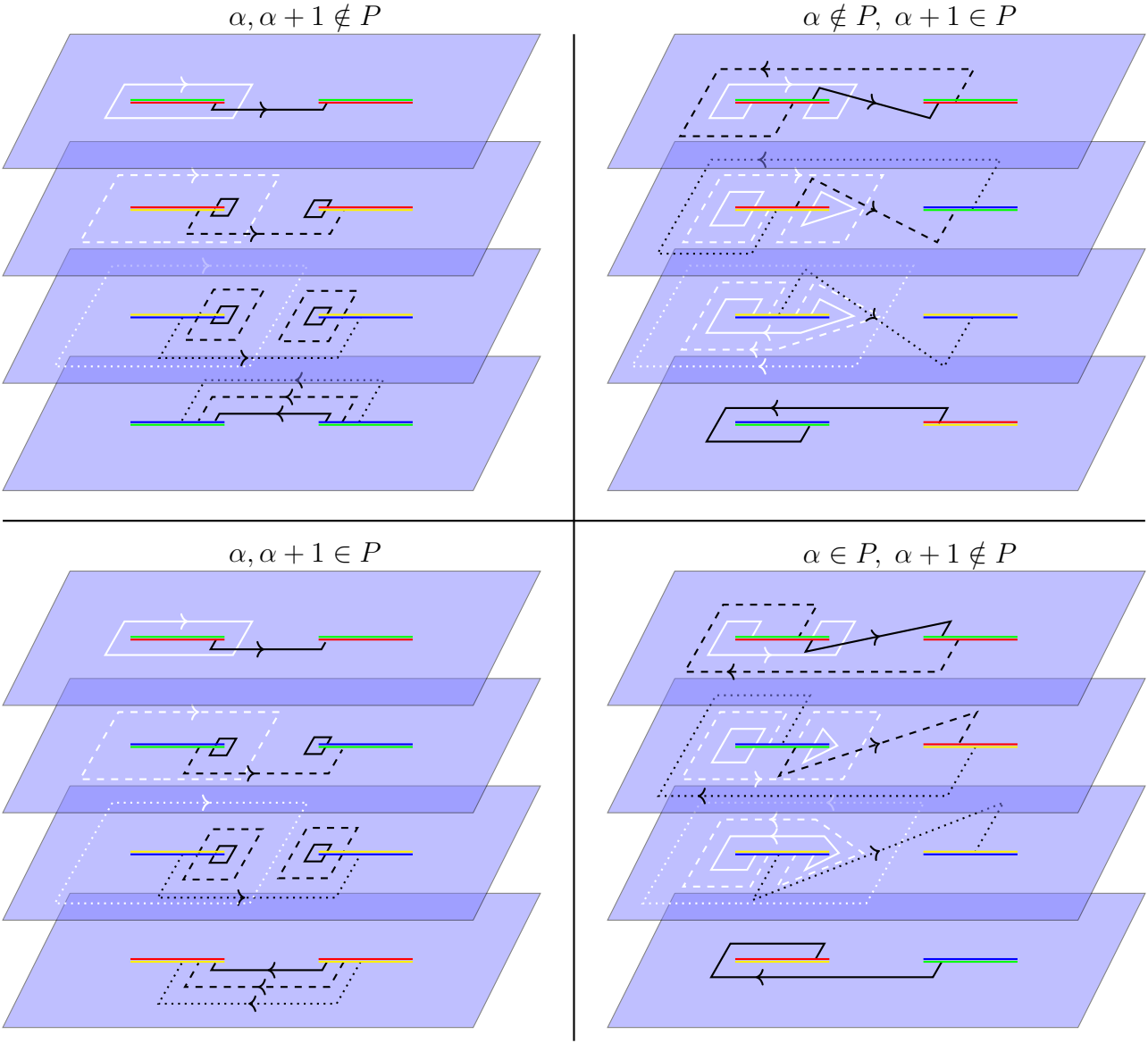

			\centering
			\pgfdeclarelayer{1}
			\pgfdeclarelayer{2}
			\pgfdeclarelayer{3}
			\pgfdeclarelayer{4}
			\pgfdeclarelayer{labels}
			\pgfsetlayers{4,3,2,1,labels}

			\caption{The basic forms for the homology loops through intervals in each of the four cases. The $a$ loops are in white and the $b$ loops are in black.}\label{fig:FourBasicLoops}
		\end{figure}

\begin{figure}[H]
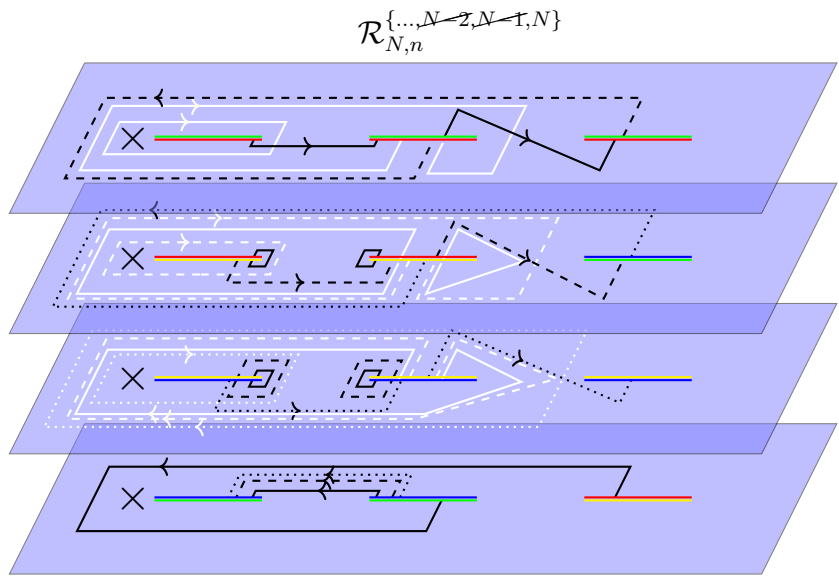

			\centering
			\pgfdeclarelayer{1}
			\pgfdeclarelayer{2}
			\pgfdeclarelayer{3}
			\pgfdeclarelayer{4}
			\pgfdeclarelayer{labels}
			\pgfsetlayers{4,3,2,1,labels}
			%
			\caption{An example for how one would construct the surface for $N\ge 3$, $n=4$ with $N-2,N-1\notin P$, $N\in P$. To construct higher $N$ surfaces, the first $N-3$ branch cuts would be inserted where the ``$\times$'' is such that the last $2$ sets of homology loops go around them.}\label{fig:GeneralNExample}
		\end{figure}

		To generalize the $N=2,P=\{2\}$ calculation to general $N$ and $P$ is fairly straightforward. The first step is to write down the four possible choices of $P$ for $N=2$: $P = \emptyset, \{1\},\{2\},\{1,2\}$. These are important as this covers the four possible relations two adjacent intervals could have: $\alpha,\alpha+1\notin P$, $\alpha\in P, \alpha+1\notin P$, $\alpha\notin P, \alpha+1\in P$, $\alpha,\alpha+1\in P$. The homology loops for these four cases are shown in figure \ref{fig:FourBasicLoops}. To construct the homology for some general $N$ and $P$, we perform the following construction: We start with the two right most intervals and draw the corresponding homology loops $a_{N-1,j}$, $b_{N-1,j}$, using figure \ref{fig:FourBasicLoops}. We then insert the $N-3$ interval (the next interval to the left), making sure to put in inside of the loops $a_{N-1,j}$, $b_{N-1,j}$. Once this is done we insert the $a_{N-2,j}$, $b_{N-2,j}$ loops, making sure to keep them inside the $a_{N-1,j}$, $b_{N-1,j}$ loops. We do this until we have drawn all the loops. An example of how this would look for $N\ge 3$ with $N\in P$, $N-1,N-2\notin P$ and $n=4$ is shown in figure \ref{fig:GeneralNExample}.

		\subsection{Calculating the Periods}
			
			The integrals $\omega_{\beta,k}(a_{\alpha,j})$ and $\omega_{\beta,k}(b_{\alpha,j})$ fall into two cases: branch cuts $\alpha$ and $\alpha+1$ have the same orientation ($\alpha\sim\alpha+1$) or they have opposite ($\alpha\nsim\alpha+1$). To evaluate them, we can use the same tricks as we used for the $N=2$ cases. The integrals can be written in terms of the line segments $c_{\alpha,j}$ and $\tilde{c}_{\alpha,j}$ where $c_{\alpha,j}$ goes from $u_{\alpha}$ to $v_{\alpha}$ on the $j^\text{th}$ sheet and $\tilde{c}_{\alpha,j}$ goes from $v_{\alpha}$ to $u_{\alpha}$ on the $j^\text{th}$ sheet. In an attempt to minimize clutter, we define $\rho_n = e^{2\pi i /n}$.
			
			\subsubsection{Periods Case 1: $\alpha\sim\alpha+1$}\hspace{\linewidth}
			
				\noindent For $\alpha\sim\alpha+1$, we are looking at the loops in the first column of figure \ref{fig:FourBasicLoops}. Note that the $b$ loops don't enclose any branch cuts, so the integrals $\omega_{\beta,k}(b_{\alpha,j})$ should not care about any other intervals besides the two they pass through. In terms of the $\tilde{c}_{\alpha,j}$s, these integrals can be written as:
				\begin{equation}
					\begin{aligned}
						\omega_{\beta,k}(b_{\alpha,j})  &= \omega_{\beta,k}(\tilde{c}_{\alpha,j}) - \omega_{\beta,k}(\tilde{c}_{\alpha,n})\\
														&= \para{\rho_n^{-k(j-1)}-\rho_n^{-k(n-1)}}\omega_{\beta,k}(\tilde{c}_{\alpha,1}).
					\end{aligned}
				\end{equation}
				For $\omega_{\beta,k}(a_{\alpha,j})$, the loops enclose branch cuts. Because of that, we need to be careful about the orientation of the cuts. So, define $c^+_{\alpha,j}$ and $c^-_{\alpha,j}$ as the paths between $u_{\alpha}$ and $v_{\alpha}$ slightly above and below the branch cuts respectively. Because we can use $T$ to change $j$, we will start with just $j=1$. We first write $\omega_{\beta,k}(a_{\alpha,1})$ as
				\begin{equation}\label{eq:omegaAsimStep1}
					\omega_{\beta,k}(a_{\alpha,1}) = \sum_{\gamma = 1}^{\alpha}\para{\omega_{\beta,k}(c^-_{\gamma,1})-\omega_{\beta,k}(c^+_{\gamma,1})}.
				\end{equation}
				Then, if $\gamma,\gamma+1\notin P$, we push $c^-_{\gamma,1}$ up through the branch cut and we have an integral of $\rho_n c^+_{\gamma,1}$. If $\gamma,\gamma+1\in P$, then we push $c^+_{\gamma,1}$ down through the branch cut to get an integral of $\rho_nc^-_{\gamma,1}$. In terms of the contributions to equation \eqref{eq:omegaAsimStep1}, the two cases differ by a minus sign that can be encapsulated in a factor $(-1)^{\delta_{\gamma\in P}}$ giving
				\begin{equation}
					\omega_{\beta,k}(a_{\alpha,1}) = \sum_{\gamma = 1}^{\alpha}(-1)^{\delta_{\gamma\in P}}\para{1-\rho_n^{-k}}\omega_{\beta,k}(c_{\gamma,1}).
				\end{equation}
				We then generate the rest of the $\omega_{\beta,k}(a_{\alpha,j})$ integrals by using $\omega_{\beta,k}(a_{\alpha,j}) = T^{j-1}\omega_{\beta,k}(a_{\alpha,1})$.
			\subsubsection{Periods Case 2: $\alpha\nsim\alpha+1$}\hspace{\linewidth}
			
				\noindent For $\alpha\nsim\alpha+1$ (second column of figure \ref{fig:FourBasicLoops}), there will be sums over contributions from the branch cuts before $\alpha$. These contributions will be the same as their contributions in the $\alpha\sim\alpha+1$ cases. It is only the contributions the final branches of the loops that change. For $\omega_{\beta,k}(b_{\alpha,j})$, there are two main components: a sum over $(\text{phases})\times(-1)^{\delta_{\alpha\in P}}\omega_{\beta,k}(-c_{\gamma,j-1})$ and $\omega_{\beta,k}(\tilde{c}_{\alpha,j} - \tilde{c}_{\alpha,j-1})$. The factor of $(-1)^{\delta_{\alpha\in P}}$ arises because the loops go along the bottom of the branch cut when $\alpha\in P$ for the $\alpha\nsim\alpha+1$ cases (see bottom right of figure \ref{fig:FourBasicLoops}).
				\begin{equation}
					\begin{aligned}
						\omega_{\beta,k}(b_{\alpha,j}) &= \rho_n^{-k(j-1)}\para{1-\rho_n^{k}}(-1)^{\delta_{\alpha\in P}}\sum_{\gamma = 1}^{\alpha}(-1)^{\delta_{\gamma\in P}}\omega_{\beta,k}(c_{\gamma,1})+\para{\rho_n^{-k(j-1)}-\rho_n^{-k(j-2)}}\omega_{\beta,k}(\tilde{c}_{\alpha,1})\\
						&= \rho_n^{-k(j-1)}\para{1-\rho_n^{k}}\para{\omega_{\beta,k}(\tilde{c}_{\alpha,1})+ (-1)^{\delta_{\alpha\in P}}\sum\limits_{\gamma = 1}^\alpha(-1)^{\delta_{\gamma\in P}}\omega_{\beta,k}(c_{\gamma,1})}
					\end{aligned}
				\end{equation}
				For $\omega_{\beta,k}(a_{\alpha,j})$, the only change from the ``$\sim$'' case is that the loops circle through sheets $j$ to $n-1$, so there will be an additional sum over sheet:
				\begin{equation}
					\begin{aligned}
						\omega_{\beta,k}(a_{\alpha,j}) &= \sum_{l=j}^{n-1}\rho_n^{-k(j-1)}\para{1-\rho_n^{-k}}\sum_{\gamma = 1}^{\alpha}(-1)^{\delta_{\gamma\in P}}\omega_{\beta,k}(c_{\gamma,1})\\
						&=\para{1-\rho_n^{-k}}\frac{\rho_n^{-kj}-1}{\rho_n^{-k}\para{1-\rho_n^{-k}}}\sum_{\gamma = 1}^{\alpha}(-1)^{\delta_{\gamma\in P}}\omega_{\beta,k}(c_{\gamma,1})
					\end{aligned}
				\end{equation}
				All of these cases are summarized in equation \eqref{eq:AllTheLoopIntegrals}.
			\begin{equation}\label{eq:AllTheLoopIntegrals}
				\begin{aligned}
					\A_{\beta,\alpha}^{k,j}=\omega_{\beta,k}(a_{\alpha,j}) &= \begin{cases} \rho_n^{-k(j-1)}\sum\limits_{\gamma = 1}^{\alpha}(-1)^{\delta_{\gamma\in P}}\para{1-\rho_n^{-k}}\int\limits_{u_{\gamma}}^{v_{\gamma}}\omega_{\beta,k}; & \alpha\sim\alpha+1\\
					\para{1-\rho_n^{-k}}\frac{\rho_n^{-kj}-1}{\rho_n^{-k}\para{1-\rho_n^{-k}}}\sum\limits_{\gamma=1}^{\alpha}(-1)^{\delta_{\gamma\in P}}\int\limits_{u_{\gamma}}^{v_{\gamma}}\omega_{\beta,k}; & \alpha\nsim\alpha+1
					\end{cases}\\\\
					\B_{\beta,\alpha}^{k,j}=\omega_{\beta,k}(b_{\alpha,j}) &= \begin{cases}\para{\rho_n^{-k(j-1)} - \rho_n^{-k(n-1)}}\int\limits_{u_{\alpha}}^{v_{\alpha+1}}\omega_{\beta,k}; & a\sim a+1\\
					\rho_n^{-k(j-1)}\para{1-\rho_n^{k}}\para{\int\limits_{u_\alpha}^{v_{\alpha+1}}\omega_{\beta,k}+(-1)^{\delta_{a\in P}}\sum\limits_{\gamma = 1}^{\alpha}(-1)^{\delta_{\gamma\in P}}\int\limits_{u_\gamma}^{v_\gamma}\omega_{\beta,k}};	& \alpha\nsim\alpha+1
					\end{cases}
				\end{aligned}
			\end{equation}
			We can now introduce the new indices $r = \beta+(N-1)(k-1)$ and $s = \alpha+(N-1)(j-1)$ to get $\tau_{rs} = (\A^{-1}\B)_{rs}$.

		\subsection{Some Remarks about the deformation $N\to N-1$}\label{sec:ChangeN}
\begin{figure}[H]
				\centering
				\begin{tikzpicture}[scale = 0.75]
					\tikzmath{\ithick = 0.1; \dis = 3; \doubledist = 0.25;}
					\draw[Gray] (-6,0)--(0,0);
					\draw[Gray] (1,0)--(7,0);
					\draw[Gray] (8,0)--(14,0);

					\filldraw[Cyan] (-5.5,-\ithick) rectangle (-5.5,\ithick);
					\filldraw[Blue] (-3.5,-\ithick) rectangle (-2.5,\ithick);
					\filldraw[Blue] (-1.5,-\ithick) rectangle (-0.5,\ithick);

					\filldraw[Cyan] (1.5,-\ithick) rectangle (2.5,\ithick);
					\filldraw[Blue] (3.5,-\ithick) rectangle (4.5,\ithick);
					\filldraw[Blue] (5.5,-\ithick) rectangle (6.5,\ithick);

					\filldraw[Cyan] (8.5,-\ithick) rectangle (10.5,\ithick);
					\filldraw[Blue] (10.5,-\ithick) rectangle (11.5,\ithick);
					\filldraw[Blue] (12.5,-\ithick) rectangle (13.5,\ithick);

					\node[below] at (3,-\ithick) {$d$};
					\node[below] at (2,-\ithick) {$l$};

					\node[above] at (4,\dis/3) {$\Tr\para{\rho_A}^{n}\brac{A_1,A_2,A_3}$};

					\draw[->,Black,thick] (0.75,0)--(0.25,0);
					\node[above] at (-3,\dis/3) {$\lim\limits_{l\to0}\Tr\para{\rho_A}^{n}\brac{A_1,A_2,A_3}$};

					\draw[->,Black,thick] (7.25,0)--(7.75,0);
					\node[above] at (11,\dis/3) {$\lim\limits_{d\to0}\Tr\para{\rho_A}^{n}\brac{A_1,A_2,A_3}$};

					\draw[Gray] (-6,-\dis)--(0,-\dis);
					\filldraw[Blue] (-3.5,-\ithick-\dis) rectangle (-2.5,\ithick-\dis);
					\filldraw[Blue] (-1.5,-\ithick-\dis) rectangle (-0.5,\ithick-\dis);
					\draw[black, thick] (-3-\doubledist/2,-\dis/3)--(-3-\doubledist/2,-2*\dis/3);
					\draw[black, thick] (-3+\doubledist/2,-\dis/3)--(-3+\doubledist/2,-2*\dis/3);
					\node[right] at (-3+\doubledist/2,-\dis/2) {$?$};
					\node[below] at (-3,-\dis-\dis/3) {$\Tr\para{\rho_A}^{n}\brac{A_2,A_3}$};

					\draw[Gray] (8,-\dis)--(14,-\dis);
					\filldraw[Blue] (8.5,-\ithick-\dis) rectangle (11.5,\ithick-\dis);
					\filldraw[Blue] (12.5,-\ithick-\dis) rectangle (13.5,\ithick-\dis);
					\draw[black, thick] (11-\doubledist/2,-\dis/3)--(11-\doubledist/2,-2*\dis/3);
					\draw[black, thick] (11+\doubledist/2,-\dis/3)--(11+\doubledist/2,-2*\dis/3); 
					\node[right] at (11+\doubledist/2,-\dis/2) {$?$};
					\node[below] at (11,-\dis-\dis/3) {$\Tr\para{\rho_A}^{n}\brac{A^{d\to0}_1\cup A_2,A_3}$};

					\node at (10,-3.2*\dis) {\includegraphics[scale = 0.6]{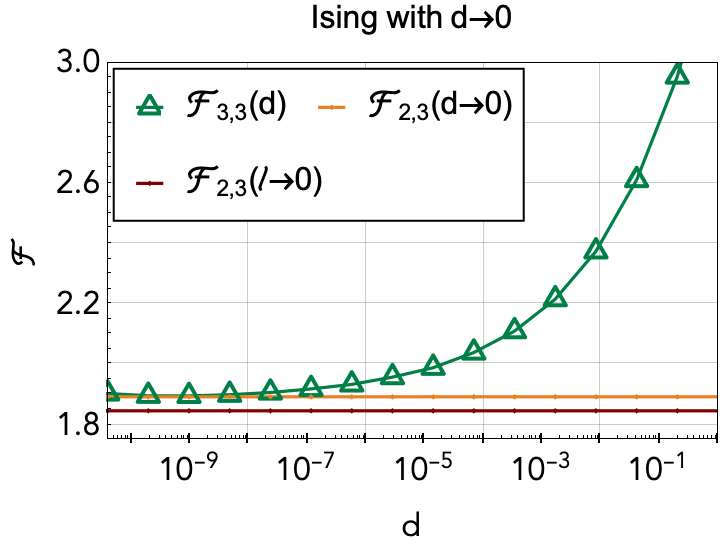}};
					\node at (-2,-3.2*\dis) {\includegraphics[scale = 0.6]{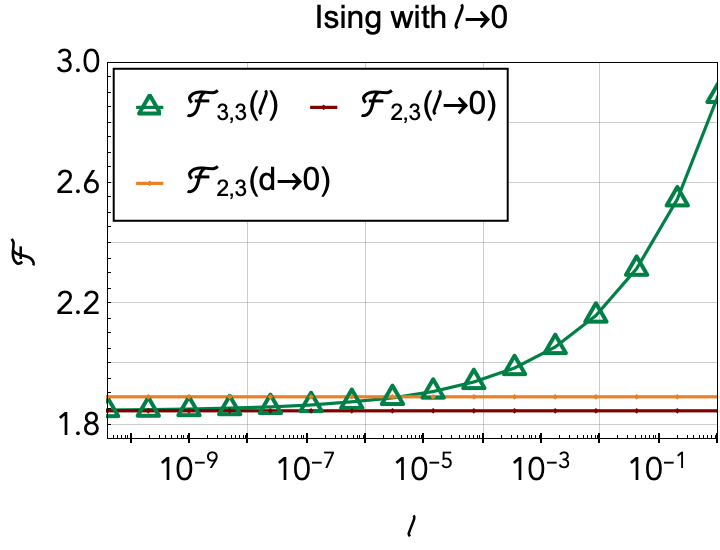}};
					\node at (4,-6*\dis) {\includegraphics[scale = 0.6]{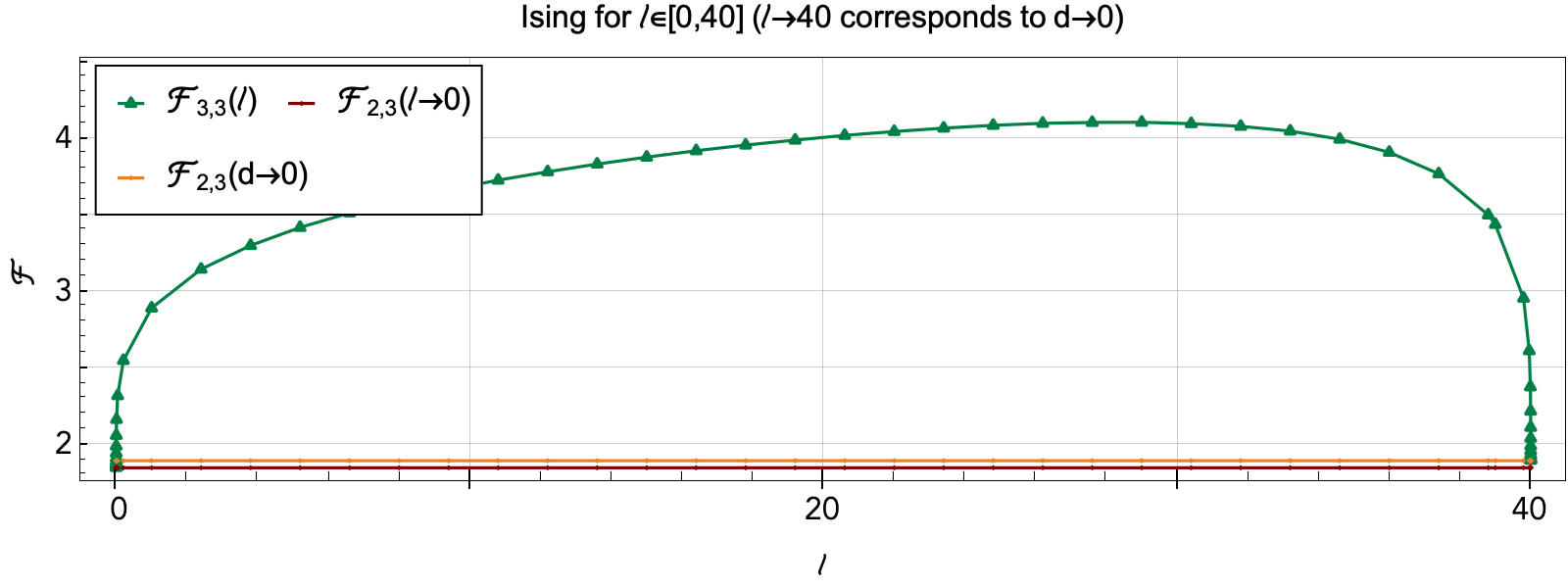}};

				\end{tikzpicture}
				\caption{We are interested in the behavior of $\Tr\para{\rho_A}^{n}\brac{A_1,A_2,A_3}$ for limits of $A_1$ where $A_1$ is stretched to touch $A_2$ and shrunk to zero.  Here, we are treating $\Tr(\rho_A)^n$ as a function of the intervals $A_1,A_2,A_3$. Because $\Tr(\rho_A)^n$ is divergent due to $\chi$, we just plot $\F$. The starting system is $A_1=\brac{u_1,v_1} = \brac{0,l_0}$, $A_2=\brac{u_2,v_2} = \brac{l_0+d_0,2l_0,d_0}$ and $A_3=\brac{u_3,v_3} = \brac{2l_0+2d_0,3l_0+2d_0}$ with $l_0=d_0 = 20$. $A_1$ is then sent to $\brac{0,0}$ ($l\to0$ limit) and $\brac{0,d_0+l_0}$ ($d\to0$ limit).}\label{fig:deformationExample1}
			\end{figure}
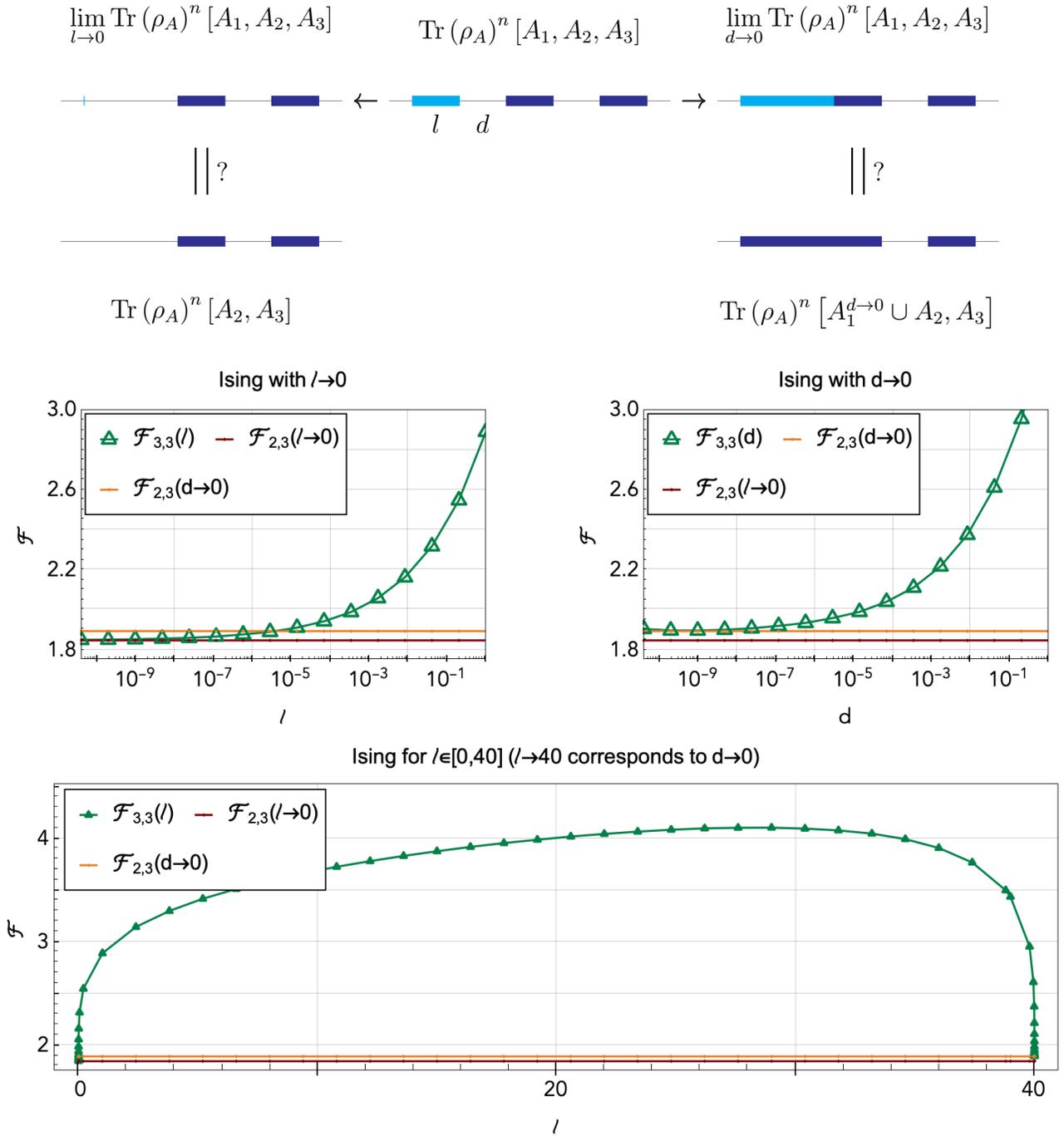

			Now that we can calculate these twist field correlation functions for general $N$, $n$ and $P$, it is natural to consider what happens when we deform the intervals. Specifically, in this section, we want to understand the behavior of $\Tr\para{\rho_A^{T_P}}^n$ in the limit where one of the intervals is shrunk to zero and if there is a function that is continuous as intervals are added or removed such as in figure \ref{fig:deformationExample1}.

			From a quick examination of equation \ref{eq:TrrhoPCFT}, its clear there is an issue as $\chi(\{u_\gamma,v_\gamma\})$ contains an overall factor of $1/(u_\mu-v_\gamma)$, which gives a divergence for both shrinking intervals to zero and bring them together. So, $\Tr\para{\rho_A^{T_P}}^n$ diverges. However, note that, for $\gamma\in P$, $(u_\gamma-v_\gamma) = -(\tilde{u}_\gamma - \tilde{v}_\gamma)$, so the ratio $R = \Tr\para{\rho_A^{T_P}}^n/\Tr\para{\rho_A}^n$ will not diverge due to this singular factor as it cancels out. The other thing to consider is $\F$. If we calculate $\F$ numerically, we see that these limits are well behaved and that $\F$ does indeed converge\footnote{Note, we say converge rather than equal, as $\F$ will be undefined at the points where $v_{\gamma-1}=u_{\gamma}$ or $u_\gamma=v_\gamma$, or in terms of figure \ref{fig:deformationExample1}, when $d$ or $l$ equal zero.} (see figure \ref{fig:deformationExample1}) except in the case that where a partially transposed and non partially transposed interval are brought together (this would correspond to replacing $\rho_A$ with $\rho_A^{T_2}$ in figure \ref{fig:deformationExample1}). This means that, save for the aforementioned case, $\F$ and $R$ are well behaved in these limits. A next step, beyond the scope of this paper, would be to look at analytic behavior of $\F$ in these limits as the Riemann theta functions diverges in the $d\to0$ limit (but the divergences cancels in $\F$) and converge in the $l\to0$ limit. 

	\section{Conclusions}
		In this paper, we have constructed a method to calculate the entanglement negativity between disjoint subsystems that are themselves made up of disjoint subsystems. This is a generalization of the calculation of entanglement negativity entropy presented in \citep{calabreseEntanglementNegativityQuantum2012}. So far we really only focused on the Ising CFT, however we mentioned that this can be used for other CFTs like the free compact boson and its decompactification regime along with others \citep{coserRenyiEntropiesDisjoint2014,dijkgraafConformalFieldTheories1988}. This is because they all just depend on the period matrix of the Riemann surface, meaning our focus on the Ising CFT does not incur a loss of generality.

		Unfortunately, with the given results, the analytic continuation of $n$ required to compute the logarithmic negativity $\E = \ln\Tr\abs{\rho_A^{T_2}}$ is not something we were able to do. The formula for $\F$ when expressed in terms of the Riemann theta functions is not algebraic in the points $u_\alpha,v_\alpha$. It is known that for $N=2$, $\RNn_{n,N}^P$ can be rewritten as a hyperelliptic curve $\nu^2 = f(\mu)$ where $f(\mu)$ is a degree $2n$ curve \citep{enolskiSingularCurvesRiemannHilbert2004,gravaEntanglementTwoDisjoint2021}. This allows $\F$ to be expressed as an algebraic function of $u_\alpha,v_\alpha$ via a Thomae formula. Unfortunately, the surfaces $\RNn_{N,n}$ and  $\RNn_{N,n}^P$ are not hyperelliptic for both $N>2$ and $n>2$, only if it is one or the other. This means the surface can not be brought into a form like  $\nu^2 = f(\mu)$ (see \ref{sec:ProofofNonHyperellipticity}). 

		In this paper we have looked at homogenous systems. It would be interesting to apply these ideas to inhomogenous systems such as those with interfaces \citep{sakaiEntanglementConformalInterfaces2008}, boundaries and defects \citep{royEntanglementEntropyCritical2021}, and in particular, systems with topological defects \citep{royEntanglementEntropyIsing2022}. Another interesting extension would be for gapped systems \citep{royQuantumSineGordonModel2021,castro-alvaredoBipartiteEntanglementEntropy2009}.

	\ack
		The author would like to thank Ananda Roy for suggesting the problem and helpful discussions. We also acknowledge discussions with Raul Arias and thank Erik Tonni for his comments.  
	\appendix
	\section{A Few Notes on Twist Fields}\label{sec:NotesTwistFields}
		In this section, we are going to talk about the equal time commutations of the twist fields and will calculate their scaling dimensions. Before doing that however, we should lay out the different ways we can think about the replica trick and the sewing conditions. If we, for a moment, just think about $\Tr\rho_A^n$, then for the replica trick, we take our theory $(\hat\C, \L)$\footnote{The first element in this tuple is the spacetime and the second is the lagrangian}, and go to a theory $\bigotimes^n(\hat\C,\L)$. The partition function of said theory is 
		\begin{equation}
			Z_{\bigotimes^n(\hat\C,\L)} = \int\prod_{j=1}^n\brac{d\phi_j}_{(\hat\C,\L)} \exp\brac{\int_{\hat\C} dzd\bar{z}\mathcal{L}[\phi_1](z,\bar{z})+\eli+\mathcal{L}[\phi_n](z,\bar{z})}.
		\end{equation}
		The introduction of twist fields is done to enforce the sewing conditions $\mathcal{C}_{N,n}:\: \phi_j(x,\tau=0^-) = \phi_{j+1}(x,\tau=0^+)$ $\forall x\in A$ with $j\in\Z_n$. This given us the path integral 
		\begin{equation}
			Z_{\bigotimes^n_{\mathcal{C}_{N,n}}(\hat\C,\L,)} = \int_{\mathcal{C}_{N,n}}\prod_{j=1}^n\brac{d\phi_j}_{(\hat\C,\L)} \exp\brac{\int_\C dzd\bar{z}\mathcal{L}[\phi_1](z,\bar{z})+\eli+\mathcal{L}[\phi_n](z,\bar{z})}.
		\end{equation}
		The subscript $\mathcal{C}_{N,n}$ represents the integral over the subspace that satisfies the sewing conditions. This theory can be thought of in two other ways: $(\hat\C,\L^n,\mathcal{C}_{N,n})$ and $(\RNn_{N,n},\L)$, the first one is considering one spacetime $\hat\C$ with $n$ copies of the fields with sewing conditions acting as identifications conditions for the fields, and the second is the theory taking place on a multisheeted Riemann surface with just one copy of the fields. From this point of view, the sewing conditions act to stich together the spacetime sheets of the replicas. It is this form of the theory that we focused on in the introduction and that will be the main focus going forward. The path integral on this spacetime is
		\begin{equation}
			Z_{(\RNn_{N,n},\L)} = \int[d\phi]_{(\RNn_{N,n},\L)}\exp\brac{-\int_{\RNn_{N,n}}dzd\bar{z}\mathcal{L}[\phi](z,\bar{z})}.
		\end{equation}

		To understand the commutation relations, let us consider $(\hat\C,\L^n)$ point of view. For $\T_n(u)$ and $\tilde\T_n(v)$ with $\Im u = \Im v$ (equal time), we have $\brac{\T_n(u),\tilde\T_n(v)} = 0$. That is, there commute at equal time. To see this, let $c_\T$ and $c_{\tilde\T}$ be counter clockwise paths that only encloses $\T$ and $\tilde\T$ respectively such that, then transporting $\phi_i$ around $c_\T$ brings you to $\phi_{i+1}$ and around $c_{\tilde\T}$ gives you $\phi_{i-1}$. With this, we can consider transport of $\phi_{i}$ around both $\T$ and $\tilde\T$. The two ways to do this are $c_\T\circ c_{\tilde\T}$ and $c_{\tilde\T}\circ c_{\T}$\footnote{The ``$\circ$'' operation on loops is composition.}. Both of these get you from $\phi_i(z)$ to $\phi_i(z)$. Thus, the order in which the loops are taken does not matter. This is a statement that $\T_n(u)\tilde\T_n(v)=\tilde\T_n(v)\T_n(u)$. In terms of the $(\RNn_{N,n},\L)$ point of view, these loops are actually paths, such that (if we parametrize from $0$ to $1$) $c_{\T}(1) = e^{2\pi i /n}c_{\T}(0)$ and  $c_{\tilde\T}(1) = e^{-2\pi i/n}c_{\tilde\T}(0)$. In other words, $c_{\T}$ starts on the $j^\text{th}$ sheet and goes to the $j+1$ sheet while $c_{\tilde\T}$ takes you to the $j-1$ sheet. Under the transport of $\phi$ along these paths, we have $\phi(z_j)\to \phi(z_{j+1})$ for $c_\T$ and $\phi(z_j)\to\phi(z_{j-1})$ for $c_{\tilde\T}$\footnote{It should be noted, that when on $\RNn$, there are actually $n$ loops $c_\T^j$ and $c_{\tilde\T}^j$ which start on the $j^\text{th}$ sheet and end on the $j+1$ sheet.}.

		If we consider just one interval with endpoints $u<v$, we have the theories $(\hat\C,\L^n)$ and $(\RNn_{N,n},\L)$. Following the notation of \citep{cardyFormFactorsBranchpoint2007,calabreseEntanglementEntropyQuantum2004}, for some field primary $\mathcal{O}$ evaluated on the $j^\text{th}$ sheet, then we have
		\begin{equation}\label{eq:HowOBehavesOnRandC}
			\braket{\mathcal{O}(w_j)\cdots}_{(\RNn_{[u,v],n},\L)} = \frac{\braket{\T(u)\tilde{\T}(v)\mathcal{O}_j(z)\cdots}_{(\hat\C,\L^n)}}{\braket{\T(u)\tilde{\T}_n(v)}_{(\hat\C,\L^n)}}
		\end{equation}
		
		where $u<v$ are points on the real line and the $j$ in $T(w,j)$ indexes what sheet of $\RNn_{N,n}$ this is being evaluated at. The field $\mathcal{O}_j$ is then the operator from the $j^\text{th}$ copy of the theory. 

		Between the two theories: $(\hat\C,\L^n)$ and $(\RNn_{N,n},\L)$, we have two forms of the stress energy tensor. For $(\hat\C,\L^n)$, we have $T(z) = \sum_{j=1}^n T_j(z)$, with each $T_j(z)$ coming from the $j^\text{th}$ copy of $\L$ and $T(w)$ which comes from $(\RNn_{N,n},\L)$. Also, if the central charge of $\L$ is $c$, then the central charge of $\L^n$ is $nc$. We want to look at $\braket{T(w)}_{(\RNn_{\{u,v\},n},\L)}$. Consider the change of coordinates between $z$ on $\hat\C$ and $w$ on $\RNn_{\{u,v\},n}$
		\begin{equation}
			z(w) = \para{\frac{w-u}{w-v}}^{1/n}.
		\end{equation}
		Then one can write
		\begin{equation}
			\braket{T(w,j)}_{(\RNn_{\{u,v\},n},\L)} = \para{\frac{\partial z}{\partial w}}^{2}\braket{T_j(z)}_{(\hat\C,\L)} + \frac{c}{12}\{z,w\}
		\end{equation}
		where $\{z,w\}$ is the Schwarzian derivative  
		\begin{equation}
			\{z,w\} = \frac{z'''z'-\tfrac{3}{2}(z'')^2}{(z')^2}.
		\end{equation}
		Note that because of translational symmetry on $\hat\C$, $\braket{T(z)} = 0$. This then becomes
		\begin{equation}
			\braket{T(w,j)}_{(\RNn_{\{u,v\},n},\L)} = \frac{c(n^2-1)}{24n^2}\frac{(u-v)^2}{(w-u)^2(w-v)^2}.
		\end{equation}
		From equation \eqref{eq:HowOBehavesOnRandC}, we then have
		\begin{equation}\label{eq:Expectation(TtildeTStressEnergy)/Expectation(TtildeT)}
			\begin{aligned}
				\frac{\braket{\T(u)\tilde{\T}(v)T(w)}_{(\hat\C,\L^n)}}{\braket{\T(u)\tilde{\T}_n(v)}_{(\hat\C,\L^n)}} &= \sum_j\frac{\braket{\T(u)\tilde{\T}(v)T_j(w)}_{(\hat\C,\L^n)}}{\braket{\T(u)\tilde{\T}_n(v)}_{(\hat\C,\L^n)}} \\
				&= 	\sum_{j}\braket{T(w,j)}_{(\RNn_{\{u,v\},n},\L)}\\
				&=  \frac{nc(n^2-1)}{24n^2}\frac{(u-v)^2}{(w-u)^2(w-v)^2}.
			\end{aligned}
		\end{equation}
		This top term is
		\begin{equation}\label{eq:Expectation(TtildeTStressEnergy)}
			\begin{gathered}
				\braket{\T(u)_n\tilde{\T}_n(v)T(w)}_{(\hat\C,\L^n)} = \hspace{4cm}\\
				\para{\frac{1}{w-u}\frac{\partial}{\partial u} + \frac{h_n}{(w-u)^2} + \frac{1}{w-v}\frac{\partial}{\partial v} + \frac{h_n}{(w-v)^2}}\braket{\T_n(u)\tilde{\T}_n(v)}_{(\hat\C,\L^n)}
			\end{gathered}
		\end{equation}
		where $\Delta_n = h_n+\bar{h}_n$ is the scaling dimension of the twist fields and $h_n$ ($\bar{h}_n$) is the holomorphic (antiholomorphic) conformal dimension. For the twist fields, $h_n=\bar{h}_n$. The standard two point function for CFTs says
		\begin{equation}\label{eq:TwoPtFunctionTtildeT}
			\braket{\T_n(u)\tilde\T_n(v)}_{\hat\C,\L^n} = \frac{C}{|u-v|^{2\Delta_n}} = \frac{C}{(u-v)^{2h_n}(\bar{u}-\bar{v})^{2\bar{h}_n}}
		\end{equation}
		Plugging this into \eqref{eq:Expectation(TtildeTStressEnergy)} and then that into \eqref{eq:Expectation(TtildeTStressEnergy)/Expectation(TtildeT)} and solving for $\Delta_n$ in the $(\hat\C,\L^n)$ theory we get:
		\begin{equation}
			h_n = \frac{c}{24}\para{n-\frac{1}{n}}
		\end{equation}
		and 
		\begin{equation}
			\Delta_n = \frac{c}{12}\para{n-\frac{1}{n}}.
		\end{equation}
		Because the theories $(\hat\C,\L^n)$ and $(\RNn_{N,n},\L)$ are the same, these conformal and scaling dimensions are the same. 

	\section{Proof of non-hyperellipticity for both $N,n>2$}\label{sec:ProofofNonHyperellipticity}
		For this, we will swap out our Riemann sphere $\hat{\C}$ for the complex projective space $\C P^1$ (they are isomorphic) and we will just write $\RNn_{N,n}$ as the $P$ superscript is not important. We are interested in
		\begin{equation}
			\RNn_{N,n}: \hspace{1cm} w^n = \prod_{\alpha=1}^N(z-u_\alpha)(z-v_\alpha)^{n-1}
		\end{equation}
		First, three things: a Riemann surface $\RNn$ is hyperelliptic curve in $\C P^1\times \C P^1$ if it admits a fractional linear involution $\iota$ with $2g+2$ fixed points ($\iota$ is non trivial), a hyperelliptic Riemann surface is a double cover of $\C P^1$ and $\iota$ must commute with other automorphisms of $\RNn$. This last point is important as we have the $T$ automorphism for $\RNn_{N,n}$ which sends $T:w\mapsto e^{2\pi i /n}w$. Because of the commutation, $\iota$ will send orbits of $T$ to orbits, $\iota:[w]_T\mapsto [w']_T$. Since each orbit is characterized by a $z$, we can look at $\iota(z)$. This leaves two options: $\iota(z) = z$ or $\iota(z)\ne z$.

		If $\iota(z) = z$, then $\iota$ will send $w$ to some $e^{2\pi i k/n}w$. But because $\iota^2 = 1$, we must have $\iota(w) = -w$. If we look at $\RNn_{N,n}/\iota$ we have
		\begin{equation}
			w^{n/2} = \prod_{\alpha = 1}^N(z-u_\alpha)(z-v_\alpha)^{n-1}
		\end{equation}
		which is a curve of genus $g = (n/2-1)(N-1)$. Firstly, this only makes sense if $n$ is even. Second, for $\RNn_{N,n}$ to be hyperelliptic, one needs $\RNn_{N,n}/\iota \simeq \C P^1$ which only happens for $N=1$ or $n = 2$. 

		For $N\ne 1$, we can consider $\iota(z)\ne z$. A general note on fractional linear transformations over over $\C P^1$: they are the M\"obius transformatons that make up $PGL(2,\C)$ and take the form
		\begin{equation}
			f: z\mapsto \frac{az+b}{cz+d}
		\end{equation}
		and have at most two fixed points unless it is the identity (this can be seen by setting  $f(z)=z$ and solving the resulting quadratic equation). This means that there at most two $z$ fixed points, and because each $z$ corresponds to at most $n$ $w$s, there are at most $2n$ fixed points. If we set the max number of possible fixed points, $2n$, equal to $2g+2$ we arrive at the equation
		\begin{equation}
			2n = 2(n-1)(N-1)+2.
		\end{equation}
		 Simplifying this, we arrive at 
		 \begin{equation}
			 1 = N-1
		 \end{equation}
		 
		 which is only satisfied for $N=2$ (for general $n$). This means that $\RNn_{N,n}$ is only hyperelliptic for $N = 2$ with general $n$ or $n=2$ with general $N$, not for both $N,n>2$. This means that we can not perform a change of variables to arrive at an equation of the form $\nu^2 = g(\rho)$ where $g$ is some polynomial in $\rho$. Meaning the change of variables used in \citep{gravaEntanglementTwoDisjoint2021} does not  generalize.

	\bibliography{PaperToArxiv.bib}

\end{document}